\documentclass[preprint,12pt]{elsarticle}

\usepackage{amssymb}
\newcommand{\glt}{%
  \mathrel{\ooalign{\raisebox{.6ex}{$>$}\cr\raisebox{-.6ex}{$<$}}}
}
\usepackage{amsmath}
\usepackage{graphicx}
\usepackage{float}
\usepackage{subfloat}
\usepackage{caption}
\usepackage{hyperref}
\usepackage{multirow}
\usepackage{multicol}
\usepackage{subcaption}
\usepackage{svg} 
\usepackage[percent]{overpic} 
\numberwithin{figure}{section}
\journal{ }
\def\Ia{{I}_{\rm app}}
\def\Isn{{I}_{\rm sn}}
\def\Ihopf{{I}_{\rm hopf}}
\def\Ihom{{I}_{\rm hom}}
\def\Idhom{{I}_{\rm dhom}}

\begin{document}

\begin{frontmatter}



\title{Stability of Phase-locked States of Weakly Coupled Izhikevich Neurons\tnoteref{*}}
\tnotetext[*]{This work was supported by the Natural Sciences and Engineering Research Council of Canada}


\author[utoronto]{XinYe Cheng\fnref{equal}}

\author[uwaterloo]{Sue Ann Campbell\fnref{equal,cor}}
\ead{sacampbell@uwaterloo.ca}

\fntext[equal]{These authors contributed equally to this work.}
\fntext[cor]{Corresponding author.}


\affiliation[utoronto]{
    organization={University of Toronto Institute for Aerospace Studies,
                  University of Toronto},
    city={Toronto},
    state={Ontario},
    country={Canada}
}

\affiliation[uwaterloo]{
    organization={Department of Applied Mathematics and Centre for Theoretical Neuroscience,
                  University of Waterloo},
    city={Waterloo},
    state={Ontario},
    country={Canada}
}

\begin{abstract}
The Izhikevich model is a computationally efficient neuron model that can exhibit a wide range of firing patterns observed in the brain. 
However, it is a discontinuous dynamical system, which means the methods for applying weakly coupled oscillator theory developed for continuous dynamical systems cannot be applied. Therefore, the collective behavior of coupled Izhikevich model has not been fully studied. To our knowledge, we carry out the first computation of the phase model for an Izhikevich neuronal model with chemical synapses. Using the phase model we study the existence and stability of phase-locking states and how these vary with parameters. Moreover, we find an interesting connection between the stability of anti-phase states of the phase model and the bifurcation behaviour of the uncoupled Izhikevich model itself. The accuracy of the prediction of the phase model is supported by simulations of two weakly coupled Izhikevich neurons. 
\end{abstract}




\end{frontmatter}

\section{Introduction}
\label{intro}

Synchronization of the electrical activity of networks neurons is important for a variety of brain functions including coordinating spatially distant populations \cite{Varela2001-vh} and memory formation and perception \cite{Buzsaki2004-mg}.
Moreover, excessive synchronization have been associated with seizure activities in epilepsy \cite{Uhlhaas2006-op}. 

There is a long history of applying weakly coupled oscillator theory to study the synchronization of biophysical neural networks \cite{hansel1993phase,hansel1995synchrony,Erm96,hoppensteadt2012weakly,ermentrout2001effects,lewis2004dendritic,saraga2004location}. However this theory has only recently been extended to piecewise smooth ~\cite{park2018infinitesimal,wilson2019isostable,Wangetal2021} and discontinuous \cite{ladenbauer2012impact,shirasaka2017phase,chartrand2019synchronization} models.  In particular, Ladenbauer et al.~\cite{ladenbauer2012impact} extended it to the adaptive exponential integrate and fire model to study the effects of subthreshold and spike-triggered adaptation 
on phase-locked states under excitatory and inhibitory coupling.  
Further, Chartrand et al.~\cite{chartrand2019synchronization} studied electrically coupled resonate-and-fire neuronal network, deriving the associated phase model to determine how spike-mediated and subthreshold coupling affect synchronization.

The synchronization state of networks of coupled neurons has been linked to a property of individual neurons, the membrane excitability class 
\cite{Erm96,stiefel2009effects,fink2011cellularly,al2021m,manoj2026phase}. 
As described by Hodgkin \cite{hodgkin1948}, neurons with the Class I excitability begin repetitive firing with
zero frequency from the resting state.
In contrast, Class II neurons start firing with non-zero frequency from the resting state 
\cite{izhikevich2007dynamical,ermentrout2010mathematical}.   Based on these firing characteristics, it has become common to associate the two classes with different bifurcations in neural models. 
Class I is typically associated with a saddle node on an invariant circle bifurcation (SNIC), while Class II is associated with a subcritical Hopf bifurcation \cite{rinzel1998analysis,ermentrout2010mathematical, ashwin2016mathematical}.  

While some cells are present as strictly Class I or Class II. The class type and resonant properties can vary \cite{hu2002two,prescott2006nonlinear,prescott2008pyramidal,stiefel2008cholinergic,brown2009neural}. A common cause of this is neuro-modulators: chemicals released by neurons that can alter the behaviour of individual neurons and the synapses that connect them. It particular, the neurotransmitter acetylcholine is known to affect the excitability and synchronization properties \cite{hu2002two,stiefel2008cholinergic,brown2009neural}  of cortical and hippocampal pyramidal cells via the down-regulation of a slow potassium current. This down-regulation switches the model from class II to class I, causing in excitatory networks of such cells switch from synchronized to desynchronized. This effect has been reiterated in biophysical neural models that incorporate such a current \cite{ermentrout2001effects,stiefel2008cholinergic,stiefel2009effects,fink2011cellularly,al2021m}. This effect must considered be with care however, as acetylcholine also down-regulates other currents such as the calcium-dependent, afterhyperpolarization potassium current, $I_{AHP}$. Studies of computational models have shown that down-regulation of $I_{AHP}$ can also cause excitatory networks to switch from synchronized to desynchronized, while the cell model remains class I \cite{ermentrout2001effects}.

The Izhikevich model was proposed an alternative the model for neural electrical activity as it is computationally efficient, yet can exhibit a wide range of voltage spiking typically only found in conductance-based biophysical models such as Hodgkin-Huxley \cite{Izhikevich2003}.  It has been used to formulate large-scale models to study brain dynamics including spontaneous travelling waves\cite{izhikevich2008large}, neural assemblies \cite{prince2021separable,hummos2014intrinsic}, cortical seizures \cite{scott2020Inhibitory},  and hippocampal rhythms~\cite{ferguson2013experimentally,ferguson2015examining,ferguson2015network}.
Hence, there has been much interest in understanding the collective dynamics of network of Izhikevich models. 
By changing a parameter in the Izhikevich model,  the model can be switched from a class 1 to class 2 oscillator \cite{izhikevich2007dynamical},  mimicking the effect of acetylcholine in some biophysical conductance-based models. 
The question then arises: does/how does this change of oscillator class translate to a change in synchronization properties?
It is the purpose of this paper to study this question.

\noindent The paper is organized as follows.
In section 2, we introduce the Izhikevich model and review its bifurcation analysis. In section 3, we develop the phase model for two weakly coupled Izhikevich neurons, following the approach of Ladenbauer et al.~\cite{ladenbauer2012impact}. We study how the variation of the parameter associated with the switch in oscillation class changes the phase model and the associated predictions about synchronization in the coupled system. Finally we give results of numerical simulations of the full model and show that the phase model gives. In section 4 we discuss our results and conclusions.


\section{Single cell model and analysis}\label{sec:singlecell}
The Izhikevich model of spiking neuron \cite{izhikevich2007dynamical} is an instance of a quadratic integrate-and-fire model with an adaptive variable of the form \cite{Gerstner_2014_neuronal}:
\begin{align}
    \dot{v} &= \left[k(v - v_r)(v - v_t) - u + I_{\text{app}}\right]/C =F^v(v,u)\label{eq:mod1} \\
    \dot{u} &= a(b(v-v_r) - u) =F^u(v,u)\label{eq:mod2}
\end{align}
where v is the membrane potential or the voltage of the neuron, and u can be interpreted as current from the ion channels of the neuron. These equations are supplemented by the resetting conditions:
\begin{equation}
    \text{if } v\geq v_{\text{peak}}, \text{ then } 
    \begin{cases}
        u &\rightarrow u + d, \\
        v &\rightarrow c. \\
    \end{cases}
    \label{eq:modreset}
\end{equation}
\noindent  
The constant $C$ (pF) is the membrane capacitance of the neuron. The $v_r$ (mV) is the neuron's resting potential and $v_t$ (mV) is the neuron's instantaneous threshold potential. The time constant for the current is $a$ (1/ms) and the $b$ (nS) is the  maximal conductance of the current. 
As a specific example, we will work with a model for a regular spiking cortical pyramidal neuron \cite{izhikevich2007dynamical}.  The corresponding parameter values are given in Table~ \ref{tab:rs_parameters}. The parameter $I_{app}$ is varied; the value determines whether the neuron is quiescent (at its resting membrane potential) or spiking. We will also study the effect of the parameter $b$ on synchronization. As we show below, this parameter determines the membrane excitability class. 

\begin{table}[htbp]
    \centering
    \caption{Parameter values for the regular-spiking cortical pyramidal neuron model.}
    \label{tab:rs_parameters}
    \begin{tabular}{ccc}
        \hline
        Parameter & Value & Units \\
        \hline
        $C$               & $100$  & $\mathrm{pF}$ \\
        $k$               & $0.7$& $\mathrm{nS\,mV^{-1}}$ \\
        $v_r$             & $-60$  & $\mathrm{mV}$ \\
        $v_t$             & $-40$  & $\mathrm{mV}$ \\
        $v_{\mathrm{peak}}$ & $35$ & $\mathrm{mV}$ \\
        $a$               & $0.03$ & $(\mathrm{ms})^{-1}$ \\
        $c$               & $-50$  & $\mathrm{mV}$ \\
        $d$               & $100$  & $\mathrm{pA}$ \\
        \hline
    \end{tabular}
    \caption{Model parameters corresponding to a regular spiking cortical pyramidal cell. Taken from \cite{izhikevich2007dynamical}.}
\end{table}

\subsection{Bifurcation analysis of the smooth subsystem}

In this section, we discuss the existence and stability of equilibrium points of  the smooth subsystem \eqref{eq:mod1}-\eqref{eq:mod2}. The analysis of the dimensionless version of this model 
can be found in \cite{touboul2008bifurcation}.  In particular, it is shown  that in the $\Ia,b$ parameter space, the system has a codimension two Bogdanov-Takens bifurcation point, which organizes the dynamics of the system. For completeness, we review these results here for the dimensional version of the model that we employ. 

The v-nullcline and u-nullcline of the system are, respectively,
\begin{align}
    u &= I_{\text{app}} + k(v - v_r)(v - v_t) \\
    u &= b(v - v_r). 
\end{align}
It is common to visualize these in the plane with $u$ as the ordinate and $v$ as the abscissa. Note that the $v$-nullcline is a parabola, which moves upward as $I_{app}$ increases, while the $u$-nullcline is a straight line with slope that varies with $b$.

The equilibrium points are given by $E_j=(v_{eq,j},u_{eq,j})$ with
\begin{align*}
    v_{\text{eq,1}} &= \frac{k(v_t + v_r) + b + \sqrt{((v_t - v_r) k + b)^2 -4kI_{\text{app}}}}{2k}, \\
    u_{\text{eq,1}} &= b (v_{\text{eq,1}} - v_r), \\
    v_{\text{eq,2}} &= \frac{k(v_t + v_r) + b - \sqrt{((v_t - v_r)k + b)^2 -4kI_{\text{app}}}}{2k}, \\
    u_{\text{eq,2}} &= b (v_{\text{eq,2}} - v_r). 
\end{align*}
Note that $v_{eq,1}>v_{eq,2}$ and $u_{eq,1}\glt  u_{eq,2}$ if $b\glt 0$.

Clearly there will be two distinct equilibrium points only  if $\Ia<I_{\text{sn}}$, where
\begin{equation}
    I_{\text{sn}} = \frac{((v_t - v_r)k + b)^2}{4k}.
\end{equation}
We discuss this point further below.

To study the stability of the equilibrium points, we assume $\Ia<\Isn$ and note that the Jacobian of the RHS evaluated at the equilibrium points is
\begin{align}
    J &= \begin{bmatrix}
        \dfrac{\eta_j}{C} & -\dfrac{1}{C} \\
        ab & -a \\
    \end{bmatrix} \label{eq:Jac}
  \end{align}
  where $\eta_j = k(2v_{eq,j} - v_r - v_t)$. 
The corresponding eigenvalues are
  \begin{align}
    \lambda_{j,\pm} &=
    \frac{\eta_j-Ca \pm \sqrt{(\eta_j-Ca)^2+4Ca(\eta_j-b)}}{2C}.
    \label{eq:evals}
\end{align}

Since $\eta_1-b=\sqrt{((v_t - v_r)k + b)^2-4k\Ia}$, it follows that $\lambda_{1,-}<0<\lambda_{1,+}$, and  $E_1$ is always a saddle. 
Since $\eta_2-b=-\sqrt{((v_t - v_r)k + b)^2 -4k\Ia}$, $E_2$ will either be a node or focus with stability determined by the sign of $\eta_2-Ca$.

When $\eta_2 = Ca$, we have
\begin{equation}
    \lambda_{2,\pm} = \pm \frac{\sqrt{Ca(Ca - b)}}{C}
\end{equation} 
Thus if $b>Ca$ these will be pure imaginary. 
It can be shown that 
$E_2$ undergoes a subcritical Andronov-Hopf bifurcation in this situation. \cite{touboul2008bifurcation}.
Note that $\eta_2=Ca$ corresponds to $I_{\rm app}=I_{\rm hopf}$ where
\begin{equation}
    I_{\text{hopf}} = 
    \frac{((v_t-v_r)k+b)^2-(b-Ca)^2}{4k}
\end{equation}
and  $E_2=(v_{\text{hopf}}, u_{\text{hopf}}) = \left(\frac{k(v_r+v_t)+Ca}{2k}, b(v_{\text{hopf}} - v_r) \right).$ 

Now consider the situation with $I_{app}=I_{sn}$. There is a unique
 equilibrium point  $E_1=E_2=(v_{\text{sn}}, u_\text{sn}) = \left(\frac{k(v_t+v_r) + b}{2k}, b(v_{\text{sn}} - v_r)\right)$, with eigenvalues
 \begin{align}
    \frac{b-Ca \pm |b-Ca|}{2C}
\end{align}
As shown in \cite{touboul2008bifurcation} this is a saddle-node bifurcation. Simple calculations show that 
the bifurcation occurs with $E_2$ a stable node if $b<Ca$ and an unstable node if $b>Ca$.

\begin{figure}[h]
    \centering
    \begin{subfigure}{0.48\textwidth}
    \begin{picture}(100,100)
    \put(20,20){\vector(0,1){120}}
    \put(10,125){$b$}
    \put(20,20){\vector(1,0){130}}
    \put(130,5){$I_{\rm app}$}
     \thicklines
     \color{blue}
     {\qbezier(40,20)(40,70)(130,70)}
     {\put(135,65){\small SN}}
     {\put(55,54){\line(3,2){70}}}
     {\put(130,100){\small Hopf}}
     {\put(55,54){\line(2,5){30}}}
     {\put(90,125){\small Homoclinic}}
     \color{black}
     \put(49,51){$\bullet$}
     \put(34,51){\small BT}
     \put(50,80){I}
     \put(75,80){II}
     \put(110,80){III}
     \put(80,40){IV}
     \end{picture}
    \caption{Schematic}
    \end{subfigure}
    \begin{subfigure}{0.48\textwidth}
        \includegraphics[width=\linewidth]{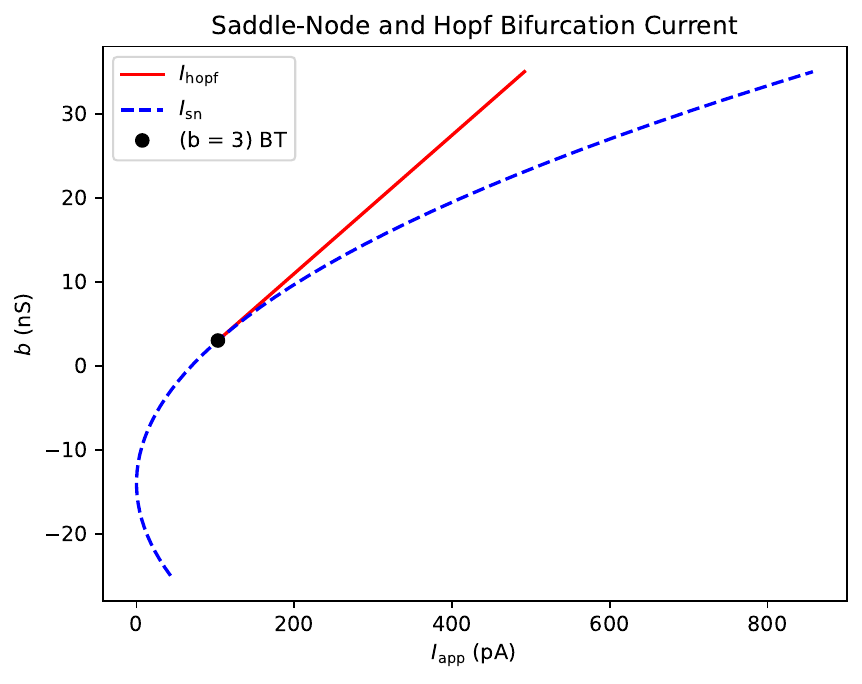}
        \caption{Actual}
    \end{subfigure}\hfill
    
    \caption{Two parameter bifurcation diagrams for the smooth subsystem \eqref{eq:mod1}-\eqref{eq:mod2}. (a) Schematic showing the general arrangement of curves for any values of the parameters. (b) Bifurcation curves for the system with parameters as in Table~\ref{tab:rs_parameters}.}
    \label{fig:2parbif}
\end{figure}

Finally, if 
$I_{\rm app}=I_{\rm sn}$ with ${b= Ca}$ there is one equilibrium point with two zero eigenvalues. Further, when $b=Ca$, we have $I_{\rm hopf}=I_{\rm sn}$. 
As shown in \cite{touboul2008bifurcation},
the point $(b,I_{\rm app})=(Ca,I_{\text{sn}})$ is a Bogdanov-Takens bifurcation point. 
Analysis of the normal form shows that in addition to the curves of saddle-node and Hopf bifurcation emanating from this point, there is a saddle-homoclinic bifurcation, which destroys the unstable periodic orbit created by the subcritical Hopf bifurcation \cite{touboul2008bifurcation}. A schematic of the bifurcation diagram is shown in Figure~\ref{fig:2parbif}(a).  The actual bifurcation curves for the parameter values of Table~\ref{tab:rs_parameters} are shown in Figure~\ref{fig:2parbif}(b). The bifurcation curves divide the parameter space into four regions with dynamics as follows
\begin{itemize}
\item[I] ($b\le Ca$ and $I_{\rm app}<I_{\rm sn}$ or $b\ge Ca$ and $I_{\rm app}<I_{\rm hom}$) 
Solutions starting on one side the stable manifold of the saddle point, $E_1$, converge to the asymptotically stable equilibrium, $E_2$. Solutions starting to the other side diverge. The structure of the stable manifold can be complicated. For model parameters in Table~\ref{tab:rs_parameters}, $b\in[-2,5]$ and $\Ia\in[0,\Isn]$ the right branch of the stable manifold lies above and fairly close to the right branch of the $v-$nullcline. For $b$ sufficiently small, the left branch lies below the $v-$nullcline for all $\Ia\in[0,\Isn]$. For larger $b$ this remains true for $\Ia$ sufficiently small, but for larger $\Ia$ the left branch can cross the $v-$nullcline and approach the right branch. See Figure~\ref{fig:manifolds}. 
\item[II] (Between the homoclinic and Hopf curves.) Solutions starting inside the unstable limit cycle converge to the stable node, $E_2$. Solutions starting outside, except those starting on the stable manifold of the saddle point, diverge.
\item[III] (Between the Hopf and SN curves). All solutions diverge except those starting on stable manifold of saddle point, $E_1$.
\item[IV] (Below/to the right of the SN curve). All solutions diverge.
\end{itemize} 

\subsection{Effect of the discontinuous reset}
The previous subsection described the dynamics of the subsystem \eqref{eq:mod1}-\eqref{eq:mod2}. Here we consider the effect of the reset \eqref{eq:modreset}. The main effect is that solutions which diverge for the subsystem \eqref{eq:mod1}-\eqref{eq:mod2} are now re-injected into the phase plane. This can lead to the existence of a large amplitude, discontinuous limit cycle. 

We consider only the situation when the parameters are as in Table~\ref{tab:rs_parameters} and focus on $b \in [-2, 5]$ as this range of $b$ values yields biologically appropriate spike frequency. With this parameter set, the BT bifurcation occurs at $b=3$ nS and $\Ia\approx 103$ pA. Further, since $c=\frac12(v_r+v_t)$ and $d=100$ the re-injection moves the solution above the $v$-nullcline. This includes the right branch of the unstable manifold of the saddle point, $E_1$, when it exists. Thus, when $E_2$ is asymptotically stable, both branches of the unstable manifold of $E_1$ can limit on $E_2$, forming a (discontinuous)  invariant loop. Further, as discussed above, for appropriate values of $b$ there exists $\Ia<\Isn$ such that the left branch of the stable manifold of $E_1$ crosses the $v$-nullcline. In this situation, it is possible there is $\Idhom<\Isn$ such that at $\Ia=\Idhom$ a homoclinic loop forms between the left stable branch and right unstable branch of $E_1$. We observe this does occur and gives rise to a stable, large amplitude discontinuous limit cycle. This bifurcation was also observed by \cite{tamura2009bifurcation}. This additional bifurcation means there is a new curve that subdivides region I into two parts, Ia and Ib. This curve intersects the SN curve at a discontinuous saddle-node homoclinic point at $b=b_{dSNH}<Ca$. The value of $b_{dSNH}$ is impossible to find exactly, instead we used a numerical approach. Using the dynamical systems package XPPAUT\cite{ermentrout_simulating_2002} we numerically approximated the stable and unstable manifolds of $E_1$ for a range of values of $b\in[0,3]$ and $\Ia$ in a range near $\Isn$. To two decimal places, we found that $b_{dSNH}\approx 1.51$. We now discuss the bifurcation sequences as $\Ia$ is increased with fixed $b$. 

\begin{figure}[hp]
    \centering
    \begin{subfigure}{\textwidth}
        \centering
         \caption{$\hspace{4in}$ }
        \vspace*{-.5in}
       \includegraphics[width=0.7\textwidth]{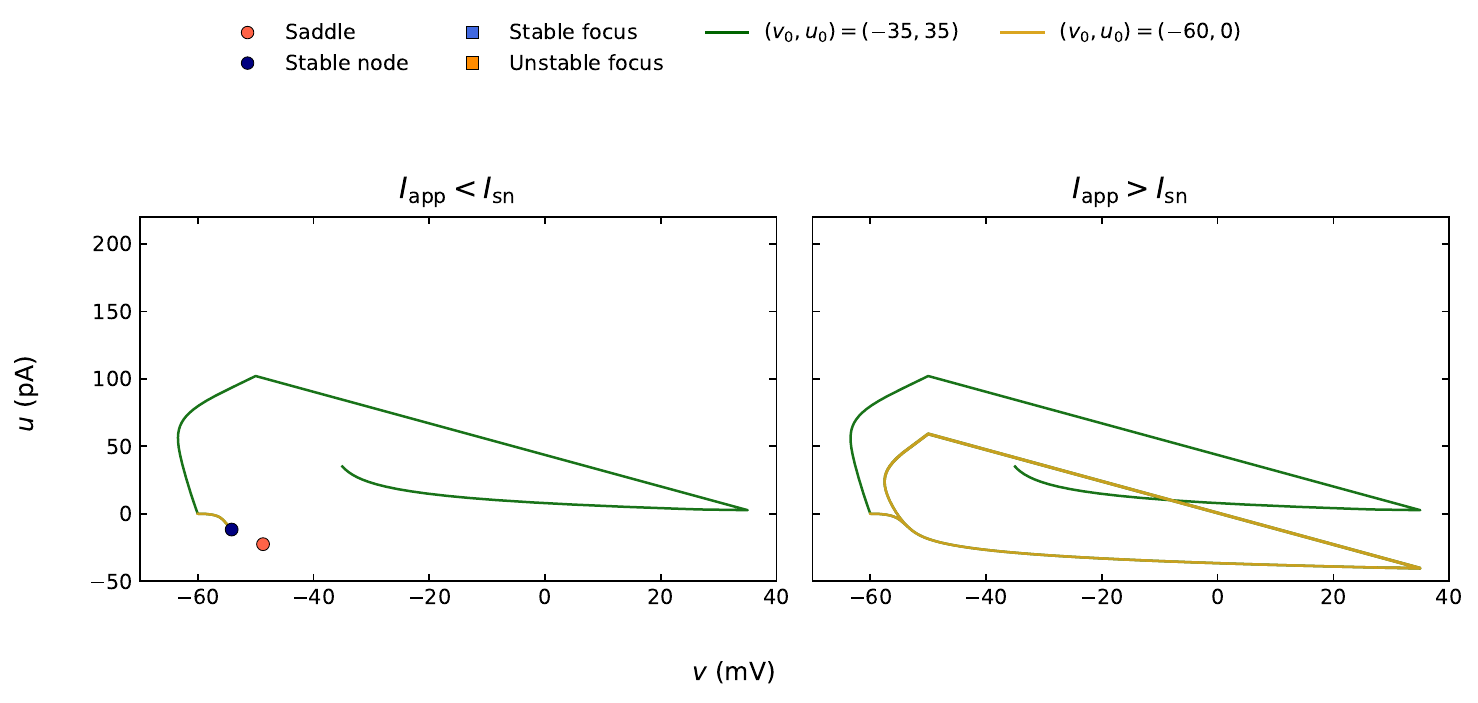}  
         \label{fig:sub-a-Bifurcation}
    \end{subfigure}
    \vskip .1in
    \begin{subfigure}{\textwidth}
       \centering
        \caption{$\hspace{4in}$}
        \includegraphics[width=0.9\textwidth]{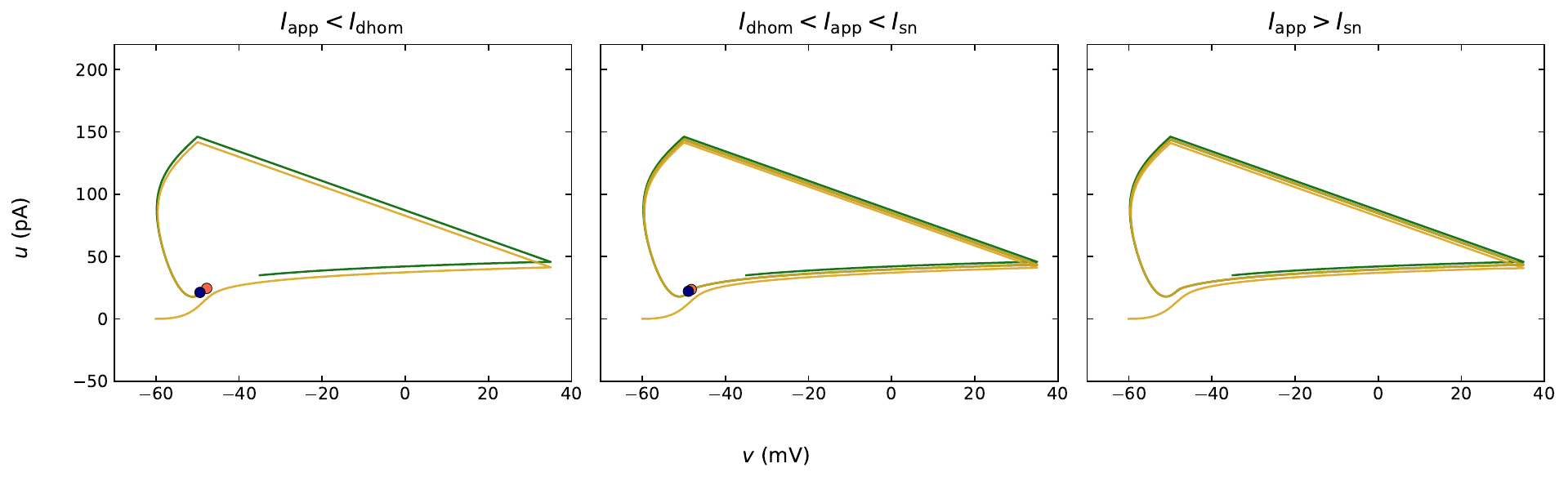} 
        \label{fig:sub-b-Bifurcation}
    \end{subfigure}
    \begin{subfigure}{\textwidth}
        \centering
         \caption{$\hspace{4in}$}
        \includegraphics[width=0.7\textwidth]{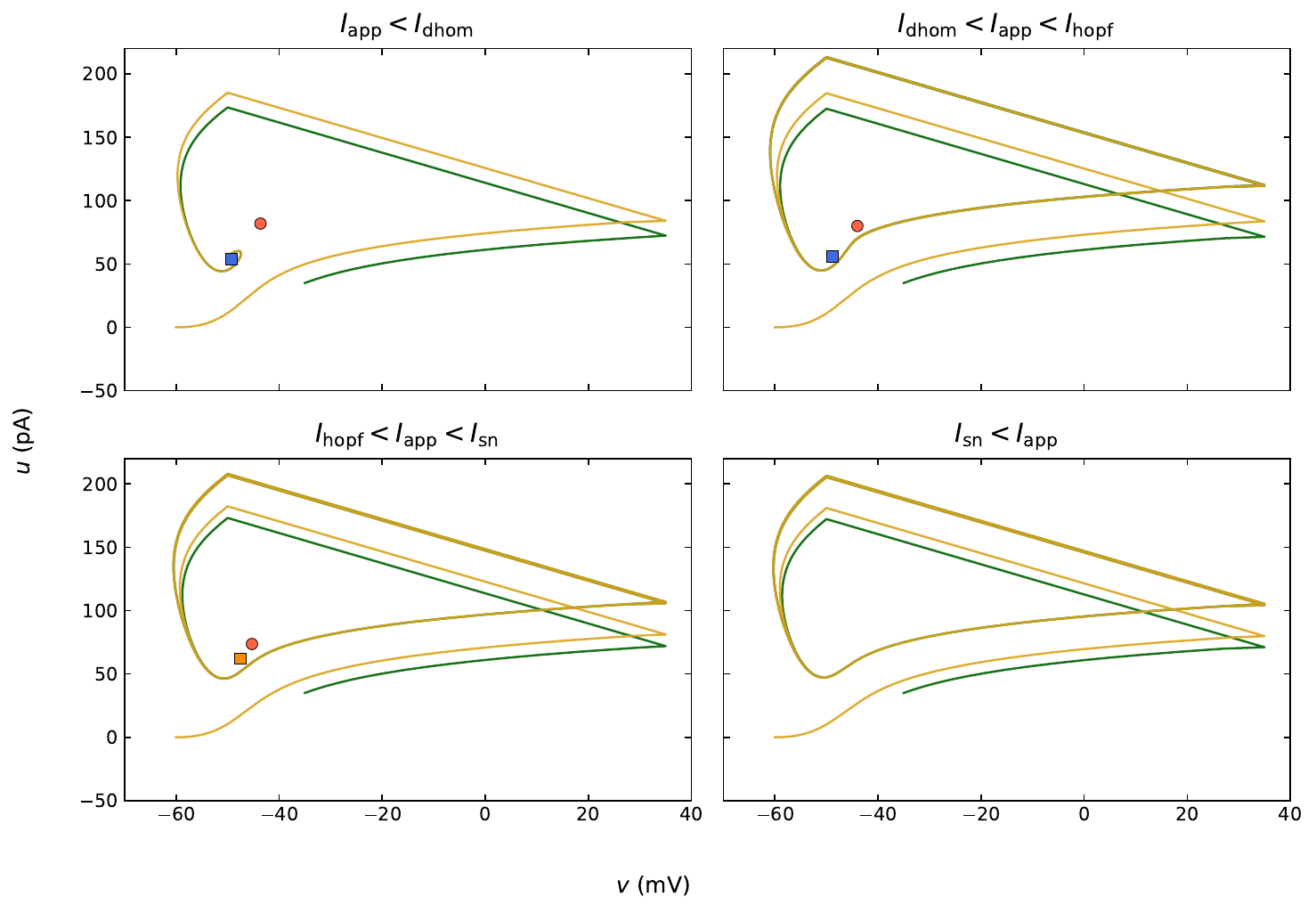} 
        \label{fig:sub-c-Bifurcation}
    \end{subfigure}
    \caption{Phase-plane simulations. (a) $b = -2$. Spiking arises in a discontinuous saddle-node on invariant circle bifurcation. Applied current values: $\Ia= 46.26, 54.00$ pA. (b) $b = 2$. Spiking arises in a saddle-node bifurcation after a discontinuous saddle-homoclinic bifurcation. Applied current values: $\Ia=90.97,91.34,91.43$ pA.
    (c) $b = 5$. Spiking arises in a subcritical Andronov-Hopf bifurcation after a discontinuous saddle-homoclinic bifurcation. Applied current values: $\Ia=123.40,124.80,128.0,129.93$ pA.
    }
    \label{fig:Bifurcation}
\end{figure}

For fixed $b$ with $b<b_{\rm dSNH}$, as $\Ia$ is increased the system transitions directly from region Ia to region IV. For $\Ia<\Isn$ the only stable object is equilibrium $E_2$ and the system is excitable. The branches of the unstable manifold of $E_1$ form an invariant loop. At $I_{\rm app}=I_{\rm sn}$ a stable node and saddle coalesce and the invariant loop becomes a stable, large amplitude discontinuous limit cycle. This is the discontinuous analog of a saddle-node on an invariant circle (SNIC) bifurcation seen in smooth models such as the Morris-Lecar model \cite{Mor_Lec}. 
This sequence is illustrated for $b = -2$ in Figure~\ref{fig:sub-a-Bifurcation}. 
When $b_{\rm dSNH}<b<Ca$, before reaching $\Isn$ the system passes through $\Ia=\Idhom$. Thus the system is excitable for $\Ia<\Idhom$, there is a region of bistability between the equilibrium $E_2$ and the large amplitude limit cycle for $\Idhom<\Ia<\Isn$, while only the stable limit cycle exists for $\Ia>\Isn$. This is illustrated for $b=2$ in Figure~\ref{fig:sub-b-Bifurcation}. Here the bifurcation at $\Ia=\Isn$ is the discontinuous analog of a saddle-node {\em off} a limit cycle discussed in \cite{izhikevich2007dynamical}.

With $b>Ca$, as $\Ia$ increases the system transitions from Region Ia through region Ib, II, III and then IV, giving the following bifurcation sequence.
For $\Ia<I_{\rm dhom}$ the only stable object is equilibrium $E_2$ and the system is excitable. At $\Ia=\Idhom$ the stable discontinuous limit cycle is created.
For $\Idhom<\Ia<\Ihom$ the stable discontinuous limit cycle and an asymptotically stable equilibrium, $E_2$, coexist. For $\Ihom<\Ia<\Ihopf$ the bistability persists, but a small amplitude unstable limit cycle surrounds $E_2$. At $\Ia=\Ihopf$ the unstable limit cycle coalesces with $E_2$ and changes its stability. For $\Ihopf<\Ia$ the only stable object is the discontinuous limit cycle. At $\Ia=\Isn$ the unstable focus and saddle point coalesce and disappear, but the stable dynamics is unchanged. 
This bifurcation sequence is illustrated for $b = 5$ \ref{fig:sub-c-Bifurcation}. It is similar to the classic subcritical Hopf sequence seen in smooth neural models such as the Hodgkin-Huxley model \cite{Hod_Hux,ermentrout2010mathematical}. 

In summary, changing the parameter $b$ changes the type of bifurcation that gives rise to stable spiking in the full model. For $b<b_{dSNH}$ the system is a Class I oscillator as the stable spiking arises in a (discontinuous) SNIC. For $b>b_{dSNH}$ the system is a Class II oscillator since when the stable equilibrium point $E_2$ disappears or loses stability the stable spiking has nonzero frequency. Although the Bogdanov-Takens point is important for organizing the behaviour of the smooth subsystem, it does not mark the transition from Class I to Class II in the full system.

\subsection{Frequency Current Curve and Excitability Class}
\begin{figure}[ht]
    \centering
    \begin{subfigure}[b]{0.45\textwidth}
        \centering
         \caption{$b = 0.2$}
        \includegraphics[width=\textwidth]{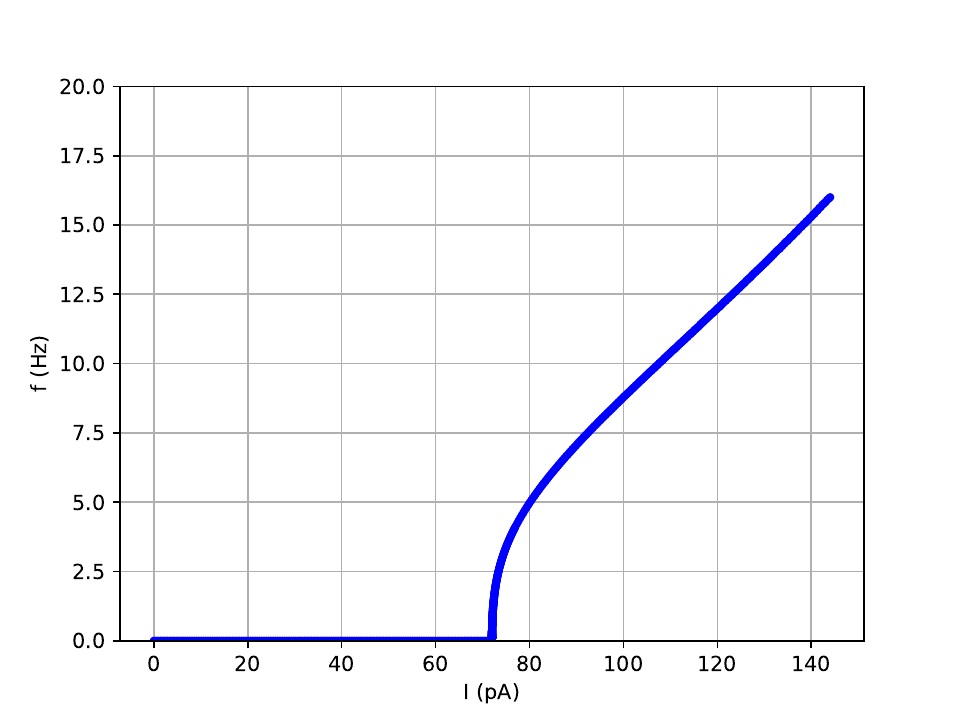}
        \vspace{-5pt}
       
        \label{fig:sub-first}
    \end{subfigure}
    \hspace{0.02\textwidth}  
    \begin{subfigure}[b]{0.45\textwidth}
        \centering
        \caption{$b = 0.6$}
    \includegraphics[width=\textwidth]{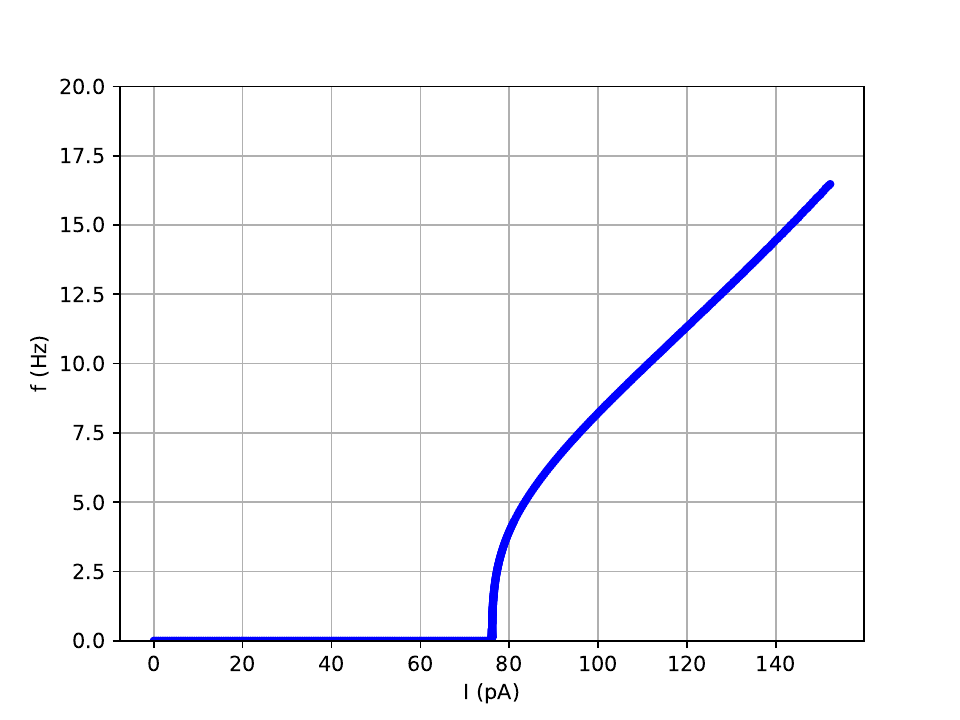}
        \vspace{-5pt}
        
        \label{fig:sub-second}
    \end{subfigure}
    
    \vfill 
    
    \begin{subfigure}[b]{0.45\textwidth}
        \centering
         \caption{$b = 2$}
        \includegraphics[width=\textwidth]{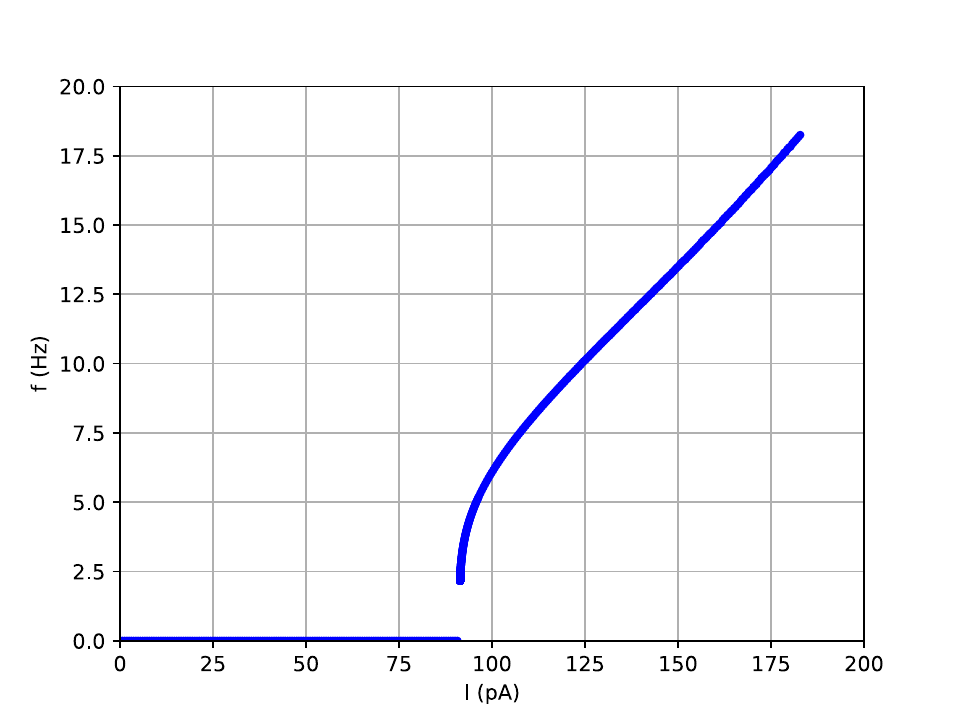}
        \vspace{-5pt}
       
        \label{fig:sub-third}
    \end{subfigure}
    \hspace{0.02\textwidth} 
    \begin{subfigure}[b]{0.45\textwidth}
        \centering
        \caption{$b = 5$}
        \includegraphics[width=\textwidth]{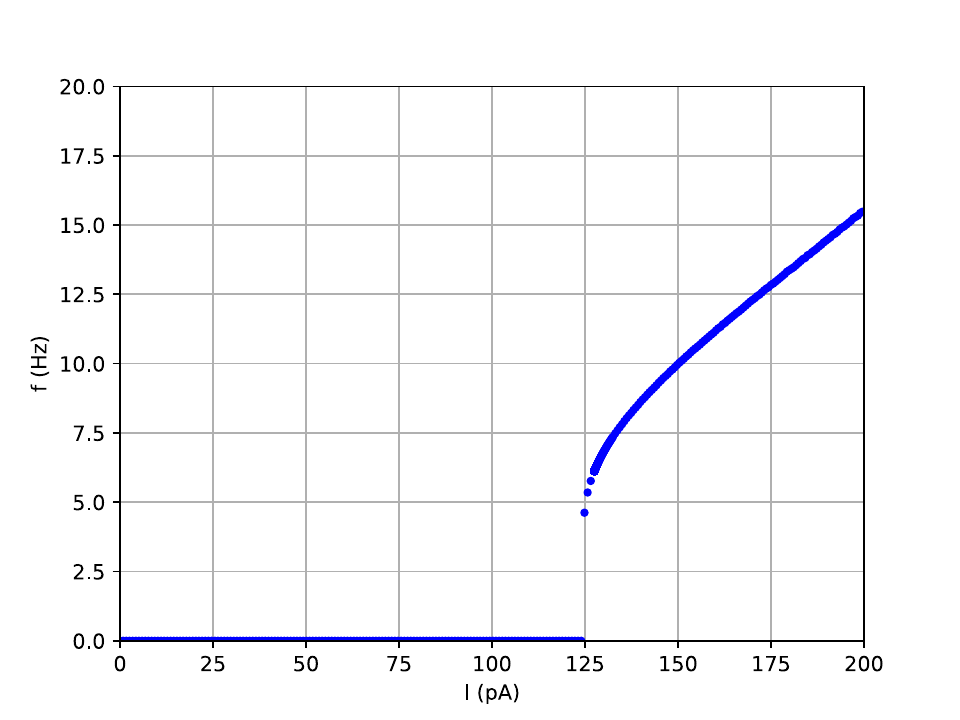}
        \vspace{-5pt}
        
        \label{fig:sub-fourth}
    \end{subfigure}
    
    \caption{F-I curves for the model with parameters as in Table~\ref{tab:Hfunresults} and $b$ values as shown.}
    \label{fig:F-I curves}
\end{figure}

The excitability class of a model can be assessed through the frequency-current (F-I) curve for the model, which shows the frequency of spiking as a function of the value of $\Ia$.
For smooth models that exhibit a BT bifurcation, it is often observed/expected that the model will exhibit Class I excitability on one side of the BT point, where the system undergoes a SNIC bifurcation and Class II on the other side, where the system undergoes a subcritical Hopf bifurcation. However, there can be ambiguity in this classification in regions where the system has bistability \cite{izhikevich2007dynamical}. To avoid ambiguity, we calculated the curves by starting the simulation very close to the stable equilibrium point when it exists. Thus our F-I curves show the onset of spiking when the equilibrium point disappears/becomes unstable. This is similar to what is observed in experiments, where $\Ia$ is typically increased with a slow ramp.   We calculated curves for $b\in[-2,5]$ with steps of $0.1$ nS. Example results are shown in Figure~\ref{fig:F-I curves}. 
We found that when $b<1.6$ the F-I curves appear continuous, indicating a class-I oscillator, while those with $b\ge 1.6$ are clearly discontinuous indicating a class-II oscillator. These results are consistent with the bifurcations described in the previous section. When the system exhibits a SNIC-like bifurcation the system is class I. When the large amplitude limit cycle is created by the discontinuous saddle-homoclinic bifurcation, spiking onset occurs via a saddle-node off the limit cycle when $b<Ca$ and via an subcritical Hopf when $b>Ca$. In either case, the limit cycle ``appears" with non-zero frequency as it already existed. 
We note that the transition point from class I to class II can be difficult to resolve accurately with numerically generated F-I curves. When the stable limit cycles are created in a discontinuous saddle-homoclinic bifurcation they have very small frequency close to this bifurcation. Near the transition point this bifurcation will be close to the saddle-node bifurcation so the ``jump" in the frequency when the stable equilibrium point disappears will be small. Nevertheless, these curves clearly show that the transition to class II occurs for $b$ close to the value predicted by our bifurcation analysis, and well before the Bogdanov-Takens bifurcation value.

\section{Phase model reduction for coupled cells}
\subsection{Background}\label{sec:phasebackground}
To begin, we review the phase model reduction for smooth systems. Consider a system of ordinary differential equation
\begin{equation}
\frac{d{\bf X}}{dt}={\bf F}({\bf X}(t)) \label{onecell}
\end{equation}
which has an exponentially asymptotically 
stable periodic orbit given by ${\bf X}=\bar {\bf X}(t), \ 0\le t\le T=2\pi/\Omega$.
Linearizing \eqref{onecell} about this solution gives the periodic system
\[\frac{d{\bf X}}{dt}=D{\bf F}(\bar{\bf X}(t)) {\bf X}.\]
The system adjoint to this is
\begin{equation}
\frac{d{\bf Z}}{dt}=-[D{\bf F}(\bar{\bf X}(t))]^T {\bf Z}.
\label{adjsys}
\end{equation}
Let ${\bf Z}=\bar{\bf Z}(t),\ 0\le t\le T$ be the unique periodic solution
of \eqref{adjsys} satisfying the normalization condition\footnote{Other conditions could be imposed. This one is commonly used in practice.}
\[ \frac{1}{T}\int_0^T \bar{\bf Z}(t)^{T}\cdot {\bf F}(\bar{\bf X}(t))\, dt=1. \]

Now consider a system of two identical coupled oscillators given by 
\begin{equation}\begin{array}{rcl}
\frac{d{\bf X}_1}{dt}&=&{\bf F}({\bf X}_1(t))+\epsilon {\bf G}({\bf X}_1,{\bf X}_2) \\[.3em]
\frac{d{\bf X}_2}{dt}&=&{\bf F}({\bf X}_2(t))+\epsilon{\bf G}({\bf X}_2,{\bf X}_1) 
\end{array}\label{genfullsys}
\end{equation}
where $\epsilon$ is the coupling strength and ${\bf G}$ is the coupling function. We assume the coupling function is the same for both connections, with the roles of the oscillators reversed.

Assuming that $\epsilon$ is sufficiently small, one can apply weakly coupled oscillator
theory \cite{EK84,EK91,HI97,ermentrout2010mathematical,schwemmer2011theory} to show that the
the phases of the oscillators of \eqref{genfullsys}  satisfy
\begin{equation}\begin{array}{rcl}
\frac{d\theta_1}{dt}&=& \Omega + \epsilon  H(\theta_2-\theta_1)
+O(\epsilon^2)\\[.3em]
\frac{d\theta_2}{dt}&=&\Omega+\epsilon  H(\theta_1-\theta_2)
+O(\epsilon^2)
\end{array}\label{phasemodel}\end{equation}
where the interaction function, $H$, is given by
\begin{equation}
H(\theta_i-\theta_j)=\frac{1}{T} \int_{0}^{T}
\bar{\bf Z}(t) {\bf G}[\bar{\bf X}(t),\bar{\bf X}(t+(\theta_j-\theta_i)/\Omega)] \, dt,
\label{Hdef}
\end{equation}
with $\hat{\bf Z},\hat{\bf X}$ as defined above. Note that $H$ is a $2\pi$
periodic function of its argument. 
Neglecting the higher order terms in $\epsilon$, introducing the
phase difference $\phi=\theta_2-\theta_1$, the system \eqref{Hdef} can be reduce to the following equation
\begin{equation}
\frac{d\phi}{dt}= H(-\phi)-H(\phi)=-H_{odd}(\phi),\label{phasediffmodel}
\end{equation}
where $H_{odd}(\phi) = H(\phi) - H(-\phi)$ is the odd part of $H$. This approach can be extended to networks of oscillators, see \cite{Erm92,ashwin1992dynamics}.

Note that equilibrium points $\phi^*$ of \eqref{phasediffmodel} correspond to the following solutions to $O(\epsilon)$ of system~\eqref{phasemodel}
\[ \begin{array}{rcl}
\theta_1(t)&=&\Omega t +\theta_{10}\\
\theta_2(t)&=&\Omega t + \phi^*-\theta_{10}.
\end{array}\label{phasesolutions}
\]
These are isolated close curves on the $2-$torus. In the original system~\eqref{genfullsys} these correspond to {\em phase-locked} periodic solutions. 
The utility of the phase model approach is that the existence and stability of the phase-locked solutions of \eqref{genfullsys} can be determined by studying the existence and stability of the equilibrium points of \eqref{phasediffmodel}. Since that is a one dimensional model, equilibrium points, $\phi^*$, are zeros of the odd part of the interaction function function, $H_{odd}$.  The stability is determined by the sign of the slope at the equilibrium point, $H'_{odd}(\phi^*)$.

Since $H_{odd}$ is $2\pi$-periodic and odd, it always admits  two equilibrium points: $\phi^*=0$ and $\phi^*=\pi$, regardless of the original model.  The first corresponds to the in-phase or synchronized solution in the full model. The second equilibrium point corresponds to the solution in the full model where the oscillators are phase-locked and are half of a period out of phase, often called the anti-phase solution. The stability of these solutions and the existence and stability of other solutions depends on the particular model being studied.
In general, the interaction function can only be computed numerically, however, the zeros and the slope can be found numerically or estimated from a plot of $H_{odd}$.

The function $\bar{\bf Z}$ is also referred to as the {\em infinitesimal phase response curve}, iPRC, and can be thought of as a measure of the change in the phase of the neuron when an arbitrarily small perturbation is applied to the voltage \cite{ermentrout2010mathematical}. The iPRC is an infinitesimal version of the PRC, which considers the effect of small but finite perturbations to a spiking neuron \cite{hansel1995synchrony}. This has been used to study synchronization properties of neurons both experimentally and computationally \cite{hansel1995synchrony,canavier1997phase,mancilla2007synchronization,bogaard2009interaction,netoff2011experimentally}. The iPRC and PRC are classified as follows \cite{hansel1995synchrony}.  Type I are always positive indicating that perturbations always tend to advance/delay the next spike. Type II have both positive and negative portions, indicating that whether the spike is advanced or delayed depends on where in the oscillation (the phase) the perturbation occurs. It has been shown that neurons with Type I membranes coupled with excitatory synapses will tend to desynchronize while those with Type II membranes will synchronize \cite{hansel1995synchrony}. The opposite is true for inhibitory coupling. ACh has been linked to a switch of PRCs from Type I to Type II both experimentally \cite{stiefel2008cholinergic} and in computational models \cite{stiefel2008cholinergic,fink2011cellularly}. The iPRC type is related to the phase model via the interaction function ~\eqref{Hdef}. We primarily work with the phase model as it gives explicit results, i.e., the exact equilibrium points and stability, which integrate the effect of coupling directly. In contrast, the iPRC (PRC) gives only qualitative results and indirectly infers the effect of coupling.

\subsection{Phase models for discontinuous dynamical systems}

In section~\ref{sec:singlecell}, 
we observe that our model does have asymptotically stable discontinuous periodic solutions. 
These can be expressed as $\bar{\bf X}(t) = (\bar{v}(t), \bar{u}(t))$ for $t \in [0, T)$, 
which satisfy \eqref{eq:mod1}-\eqref{eq:mod2} for $t\in (0,T)$ and the jump conditions
\[ \bar{v}(0)=c,\ \bar{v}(T^-)=v_{peak},\ \bar{u}(0)=\bar{u}(T^-)+d.\]
Thus the periodicity of the oscillator relies on the reset condition~\eqref{eq:modreset}. 
As described by Ladenbauer et al. \cite{ladenbauer2012impact}, the phase of such a solution is well defined and the corresponding phase model can be derived. The key adjustment is when solving the adjoint system for the iPRC, an additional boundary condition must be imposed to incorporate the effect of the reset on the iPRC. This condition ensures the iPRC is periodic. Once the iPRC is determined we can compute the interaction function, $H(\phi)$, and phase model \eqref{phasediffmodel} as in the continuous case. 
\subsubsection{Infinitesimal phase response curve (iPRC)}
Recall that the Jacobian of the continuous part of the Izhikevich model is given by equation~\eqref{eq:Jac}. 
Thus the adjoint system is 
\begin{equation}
    \begin{bmatrix} \dfrac{dZ^v}{dt} \\[12pt] \dfrac{dZ^u}{dt} \end{bmatrix} =   
    \begin{bmatrix}
    \dfrac{-k(2\bar{v} - v_t - v_r)}{C} & -ab \\[.6em]
    \dfrac{1}{C} & a
    \end{bmatrix}
    \begin{bmatrix}
       Z^v\\ Z^u 
    \end{bmatrix}
\end{equation}
 
The normalization condition we impose is 
\begin{equation}
Z^v(0)F^v(\bar{v}(0,\bar{u}(0))+Z^u(0)F^u(\bar{v}(0,\bar{u}(0))=1.
\label{eq:norm}
\end{equation}
As discussed above, since our model is discontinuous at the end of the limit cycle, to ensure the iPRC is periodic, we impose the boundary condition \cite{ladenbauer2012impact},
\begin{equation}
    Z^u(0) = Z^u(T^-).
\end{equation}
To solve the above boundary value problem in \texttt{Python}, we choose a constant $I_{\text{app}}$ large enough to produce more than 4 spikes, and take one period of the stable limit cycle, from the reset condition $v = c$ to the voltage peak $v_{\text{peak}}$. Then we use a linear interpolator \verb|numpy.interp| to convert our limit cycle, $\bar{v}$, a \texttt{NumPy} array into a function $\bar{v}(t)$. Then we pass $\bar{v}(t)$ into \verb|scipy.integrate.solve_bvp|, where we implement the boundary value problem above in \texttt{Python} and solve. 

\subsubsection{Phase model} 
Now consider the model for two Izhikevich neurons coupled via chemical synapses:
\begin{align*}
     C\dot{v}_i &= k(v_i - v_r)(v_i - v_t) - u_i + I_{\text{app}}(t)  - g_{\text{syn}} s_j(v_i - v_{\text{syn}})\\
     \dot{u_i} &= a(b(v_i - v_r) - u_i) \\
     \dot{s_i} &= -\beta s_i 
\end{align*}
where $i, j \in \{1, 2\}$, but $i \neq j$. Here $\beta > 0$ and we assume $ 0 <g_{\text{syn}} \ll 1$, to ensure weak coupling. This continuous part of the model is supplemented by the resetting conditions, 
\begin{equation}
    \text{if } v_i \geq v_{\text{peak}}, \text{ then } 
    \begin{cases}
        u_i &\rightarrow u + d \\
        v_i &\rightarrow c \\
        s_i &\rightarrow 1
    \end{cases}
\end{equation}

We use the single exponential synapse model \cite{ermentrout2010mathematical}. The gating variable $s$ represents the fraction of open channels in the postsynaptic cell membrane. When the presynaptic cell neuron fires an action potential, all the channels open up instantaneously. The channels subsequently close with time constant $1/\beta$. The constants $g_{\text{syn}}$ and $v_{\text{syn}}$ are maximal synaptic conductance (synaptic strength) and reversal potential, respectively. The synaptic input can be excitatory or inhibitory, where $v_{\text{syn}}$ is 0 mV for excitatory connection while $v_{\text{syn}}$ varies between -81 to -60 mV for inhibitory connection. Here we focus on excitatory coupling.

Note that the coupling function in our model is  ${\bf G}=(G^v,0)^T$, where $G^v(v_i, v_j) = -s_j(v_i - v_{\text{syn}})$. 
Thus, given the solution for the iPRC, $\bar{\bf Z}(t)$, the interaction function is given by 
\begin{equation}
    H(\phi) = \frac{1}{T} \int^T_0 \bar{Z}^v(t) G^v(\bar{\bf X}(t), \bar{\bf X}(t + \phi)) dt.
    \label{eq:HdefIz}
\end{equation}
For simplicity, we take $\Omega=1$, which means we measure $\phi$ in units of time.
For each value of $\phi$, this equation calculates the cumulative effect of the coupling on the postsynaptic neuron if the presynaptic neuron is out of phase an amount $\phi$.
To compute \eqref{eq:HdefIz}, we assume the presynaptic neuron fires at $t+\phi$ and hence the corresponding gating variable jumps to $1$ and subsequently follows the exponential decay equation. Then the coupling has the simple representation
\begin{equation}
   G^v(\bar{\bf X}(t), \bar{\bf X}(t + \phi)) = -\exp(-\beta (t + \phi))(\bar{v}(t) - v_{\text{syn}}). \label{eq:couplingH}
\end{equation}
Then, taking $\bar{Z}^v$ and $\bar{\bf X}$ from the previous computations,
we use \\ \verb|SciPy.integrate.simpson| to compute the integral in \eqref{eq:HdefIz}.

\section{Results}
\begin{figure}
\begin{center}
\subfloat[$b=0.2,\ I_{app}=72.03$]{\includegraphics[width=0.4\textwidth]{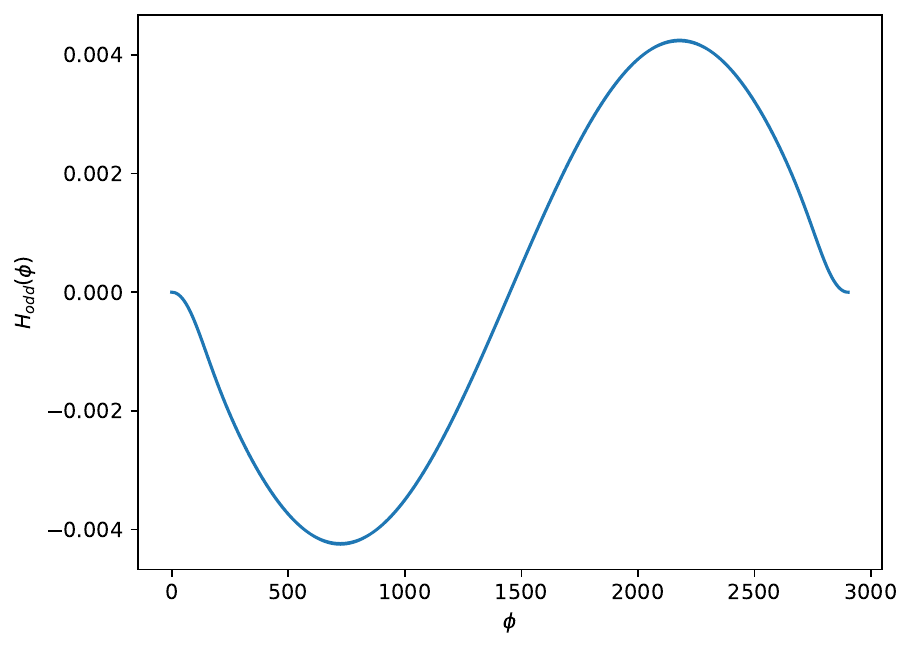}}
\subfloat[$b=0.2,\ I_{app}=72.50$]{\includegraphics[width=0.4\textwidth]{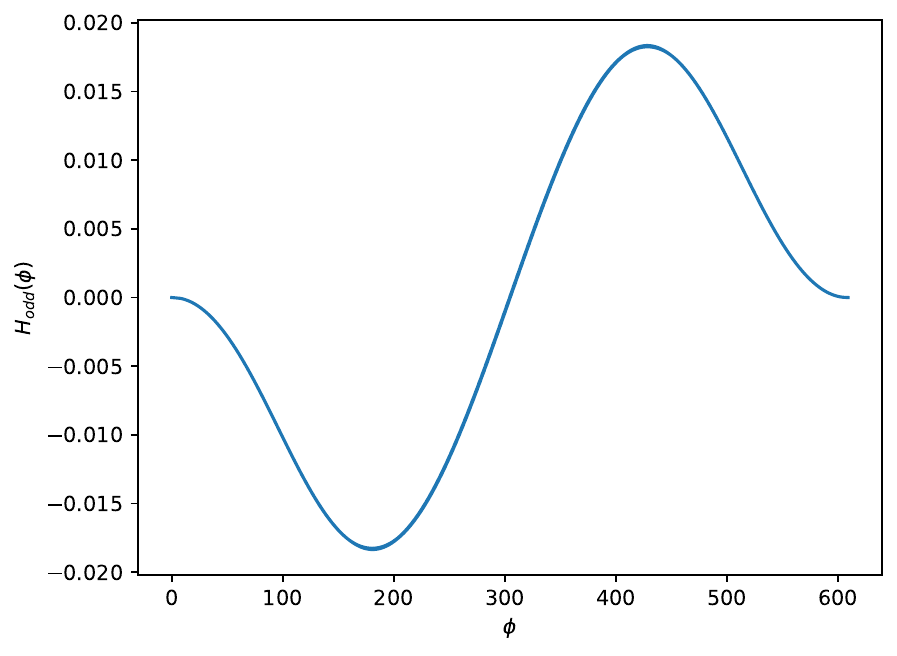}}\\
\subfloat[$b=0.6,\ I_{app}=76.14$]{\includegraphics[width=0.4\textwidth]{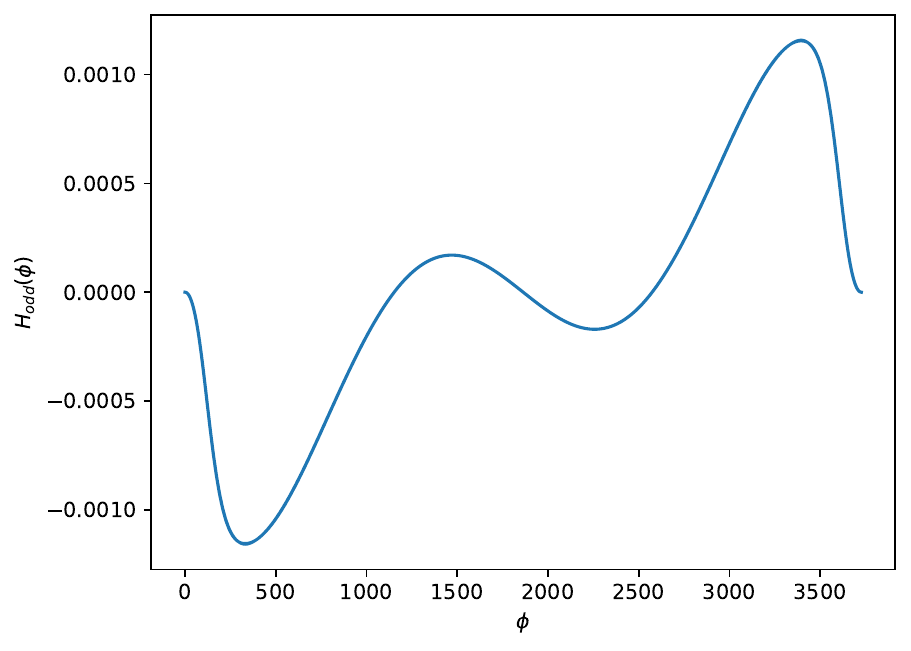}}
\subfloat[$b=0.6,\ I_{app}=77.00$]{\includegraphics[width=0.4\textwidth]{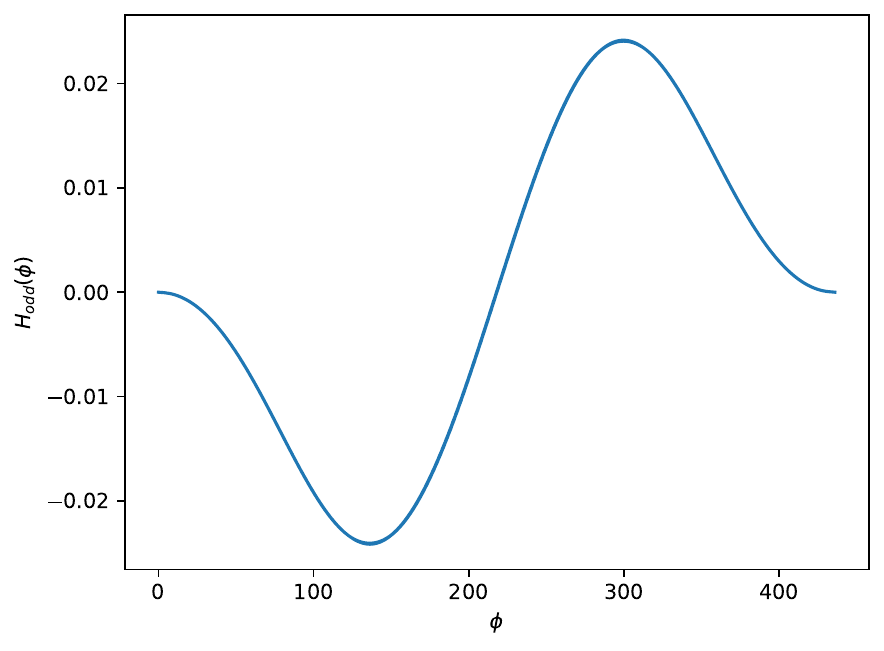}}
\\
\subfloat[$b=2,\ I_{app}=91.35$]{\includegraphics[width=0.4\textwidth]{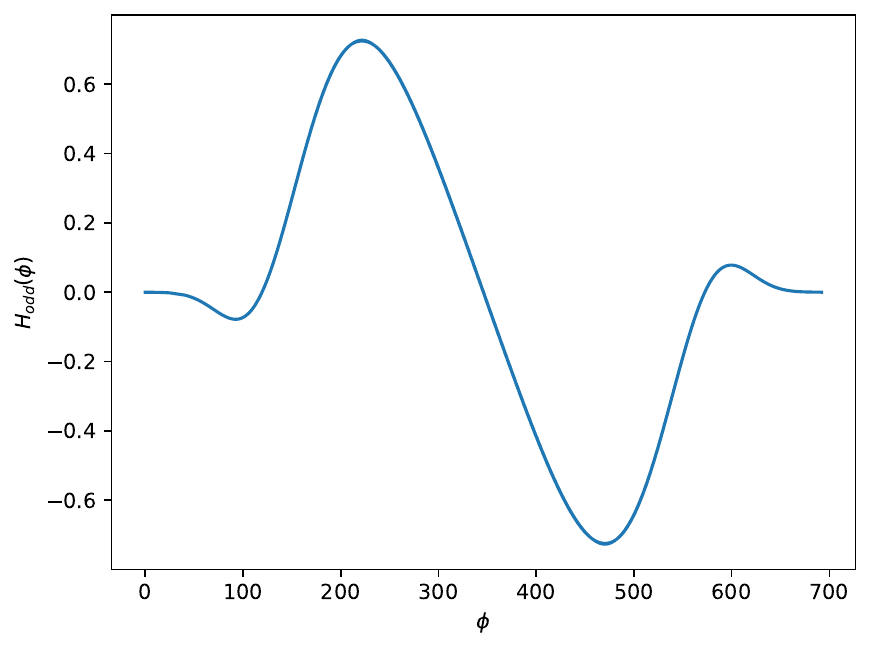}}
\subfloat[$b=2,\ I_{app}=92.34$]{\includegraphics[width=0.4\textwidth]{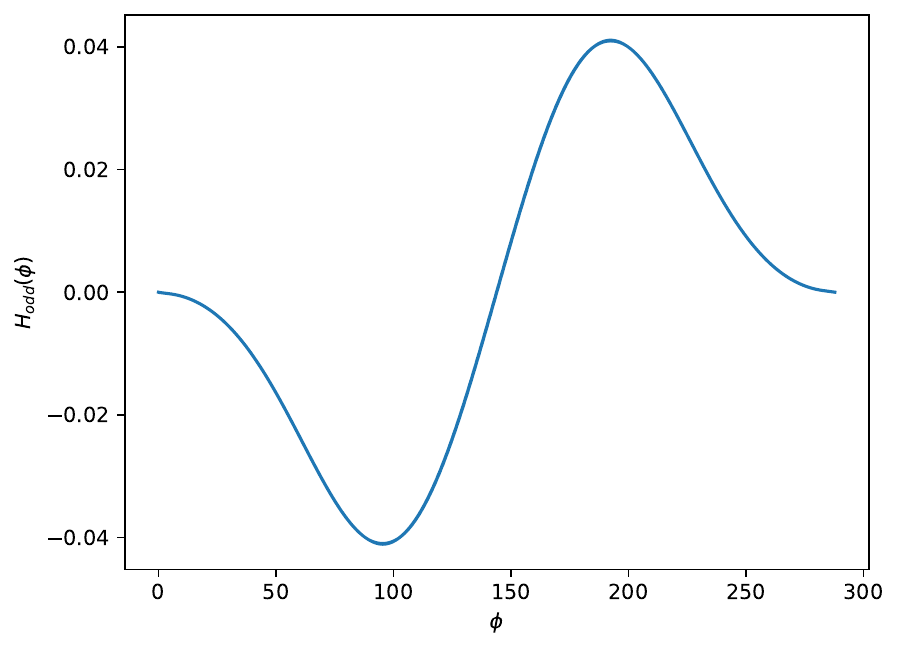}}\\
\subfloat[$b=5,\ I_{app}=124.63$]{\includegraphics[width=0.4\textwidth]{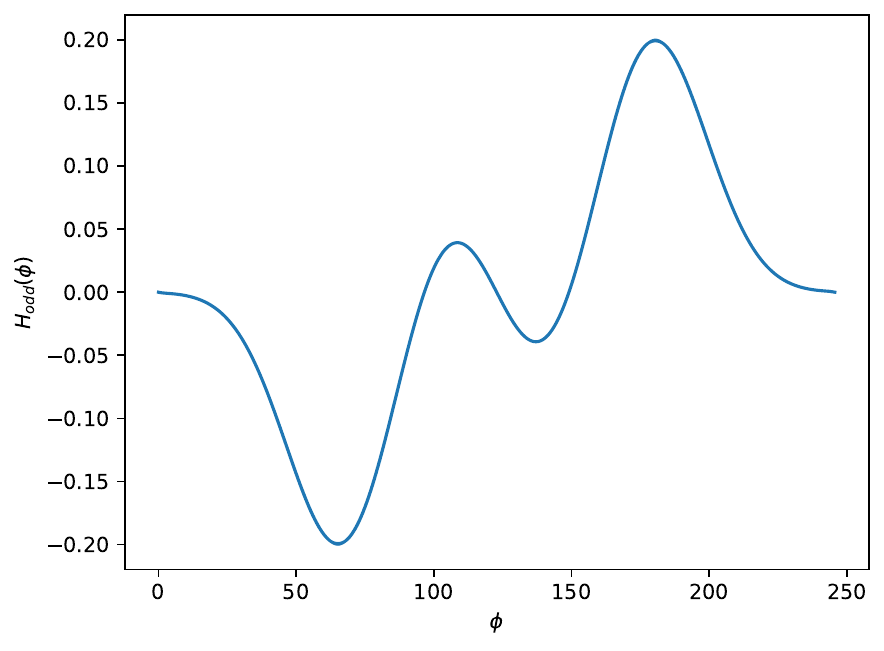}}
\subfloat[$b=5,\ I_{app}=124.95$]{\includegraphics[width=0.4\textwidth]{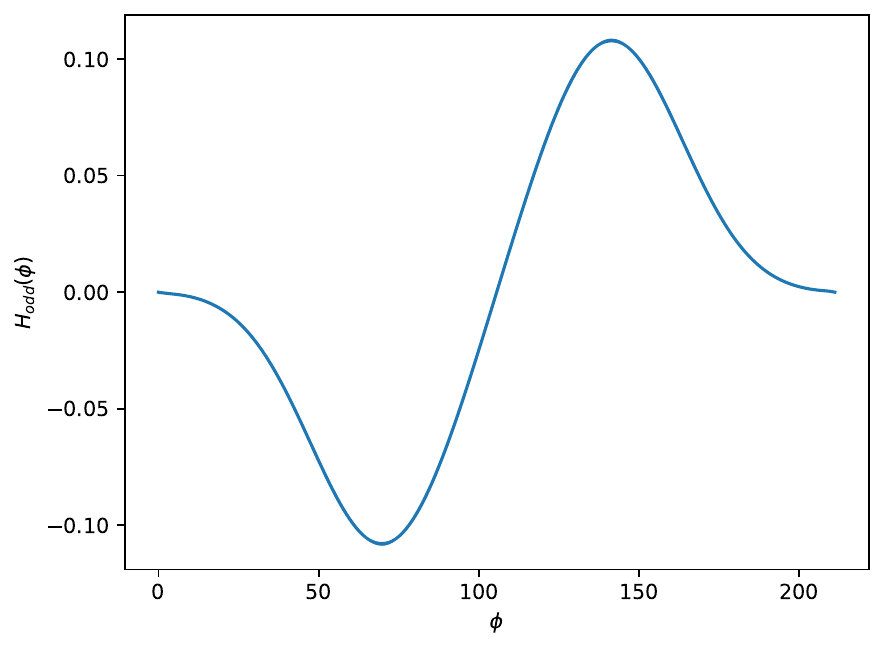}}\\
\caption{Interaction functions for values of $b$ and $I_{app}$ as shown and $\beta=0.8$. All other parameters are as given in Table~\ref{tab:rs_parameters}. The phase $\phi$ is shown in units of time (ms) over one period of the limit cycle.}
\label{fig:Hfunction}
\end{center}
\end{figure}

We calculated the interaction function, $H_{odd}$, for the phase model \eqref{phasediffmodel} as discussed in Section~\ref{sec:phasebackground} using the parameters in Table~\ref{tab:rs_parameters} and $\beta=0.8$ for a range of values $b\in [-2,5]$ and two values of $I_{app}$ for each $b$. The first $I_{app}$ value was close to the bifurcation point associated with the onset of spiking while the second was further away. The interaction function for some example parameter sets are shown in Fig.~\ref{fig:Hfunction}. Recall that these functions can be used to predict the existence and stability of phase-locked periodic solutions for a pair of coupled neurons as described in Section~\ref{sec:phasebackground}. We found two distinct cases. For $b<0.5$ only the in-phase and anti-phase solutions exist and these were predicted to be stable and unstable, respectively, regardless of the choice of $I_{app}$. See Fig.~\ref{fig:Hfunction}(a),(b). For $b\ge 0.6$, with $I_{app}$ close enough to the bifurcation point, multiple phase-locked solutions are predicted to exist with {\bf both} the in-phase and anti-phase predicted to be stable. For larger $I_{app}$ only the in phase solution was predicted to be stable. See Fig.~\ref{fig:Hfunction}(c)-(h). Note that the shape of the interaction function with $I_{app}$ close to the bifurcation point varied dramatically with $b$, while further away it was virtually unchanged with $b$. The predictions across all parameters studied are summarized in Table~\ref{tab:Hfunresults}.

To supplement our results, we varied the synaptic time constant parameter, $\beta$, and found that lower values ($\beta = 0.1, 0.2, 0.4, 0.6$) produced interaction functions with qualitatively the same shape, i.e., giving the same predictions of existence and stability of phase-locked states. Note that lower $\beta$ means that the decay rate of the synaptic channels is smaller increasing the time that the neurons influence each other. Correspondingly, lower values of $\beta$ gave interaction functions with larger magnitude predicting faster convergence to stable phase-locked states. 

Finally, we considered two other values of the jump in the current $d=20,50$. We focused on $b>0$ as with $b<0$ and sufficiently small $d$ the limit cycles become almost one dimensional, have very fast spiking and are difficult to resolve numerically. For most parameter values we considered, the interaction function was not qualitatively changed. For values of $b$ where the stability of the anti-phase solutions depends on $\Ia$, smaller $d$ allowed stable anti-phase solutions for larger $\Ia$, however the anti-phase solution was still unstable for large enough $\Ia$. This is not too surprising as decreasing $d$ increases the value of $\Idhom$ where the stable discontinuous limit cycles arise. 

\begin{figure}
\begin{center}
\subfloat[$b=0.2,\ I_{app}=72.03$. ]{\includegraphics[width=0.72\textwidth]{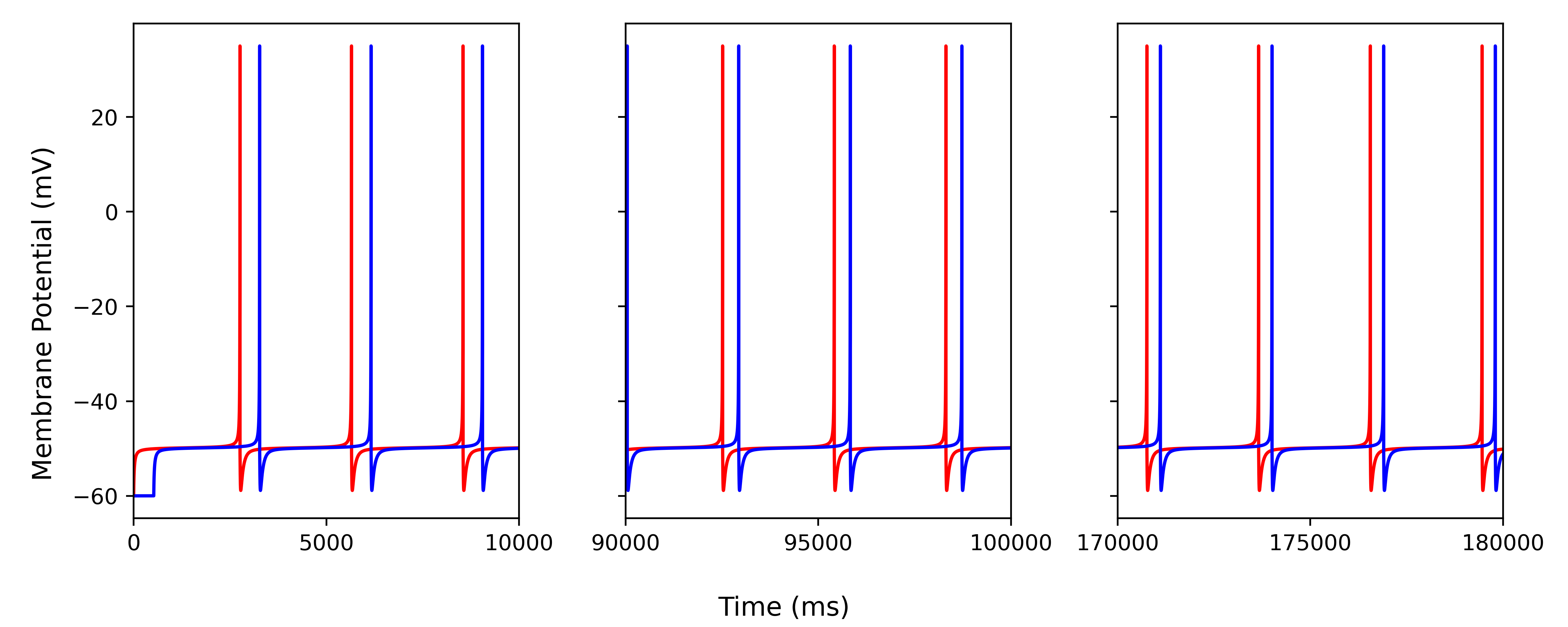}}\\
\subfloat[$b=2,\ I_{app}=91.35$]{\includegraphics[width=0.72\textwidth]{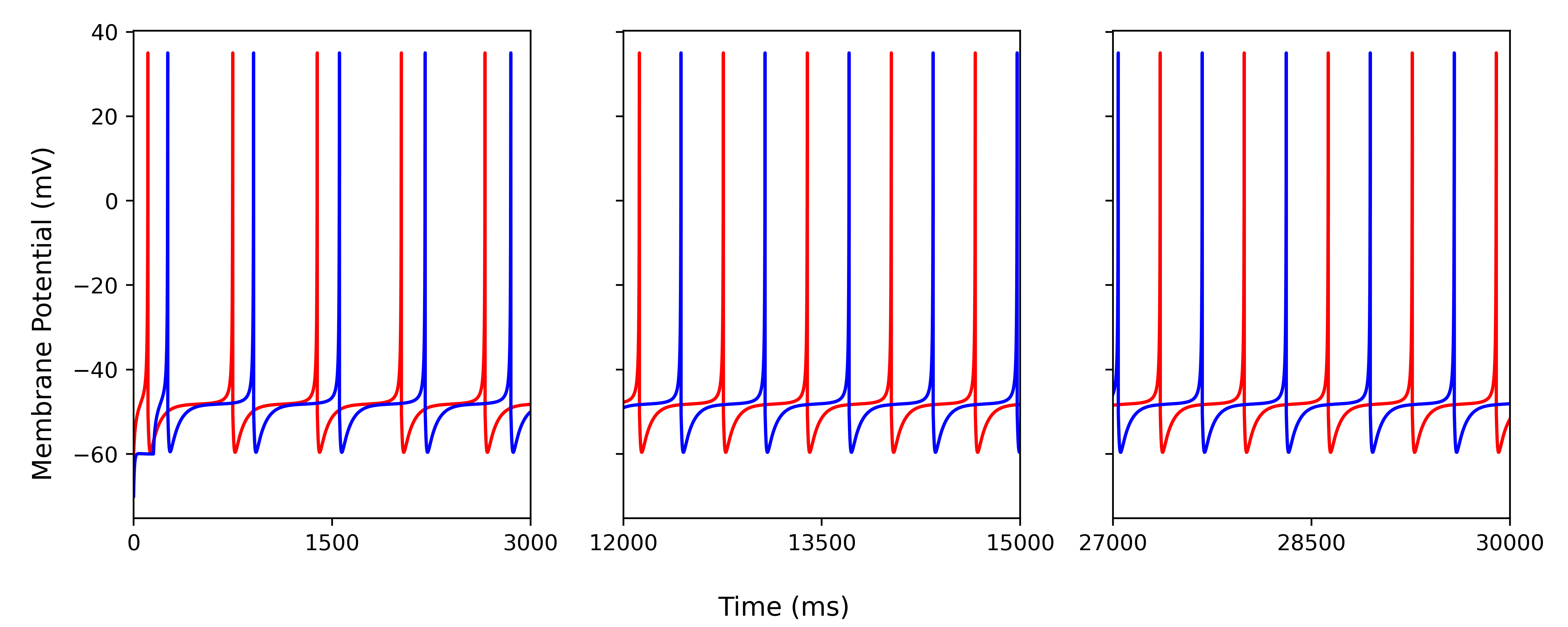}}\\
\subfloat[$b=2,\ I_{app}=91.35$]{\includegraphics[width=0.72\textwidth]{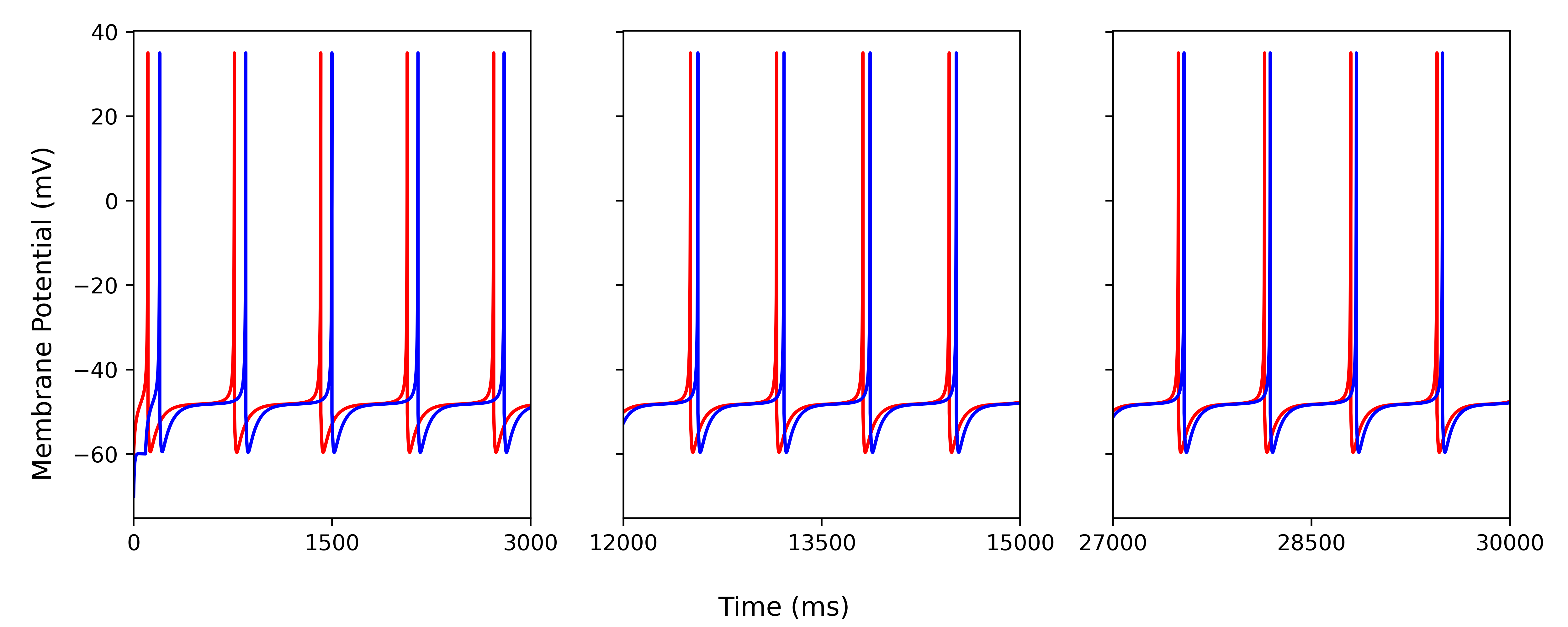}}\\
\subfloat[$b=2,\ I_{app}=92.34$]{\includegraphics[width=0.72\textwidth]{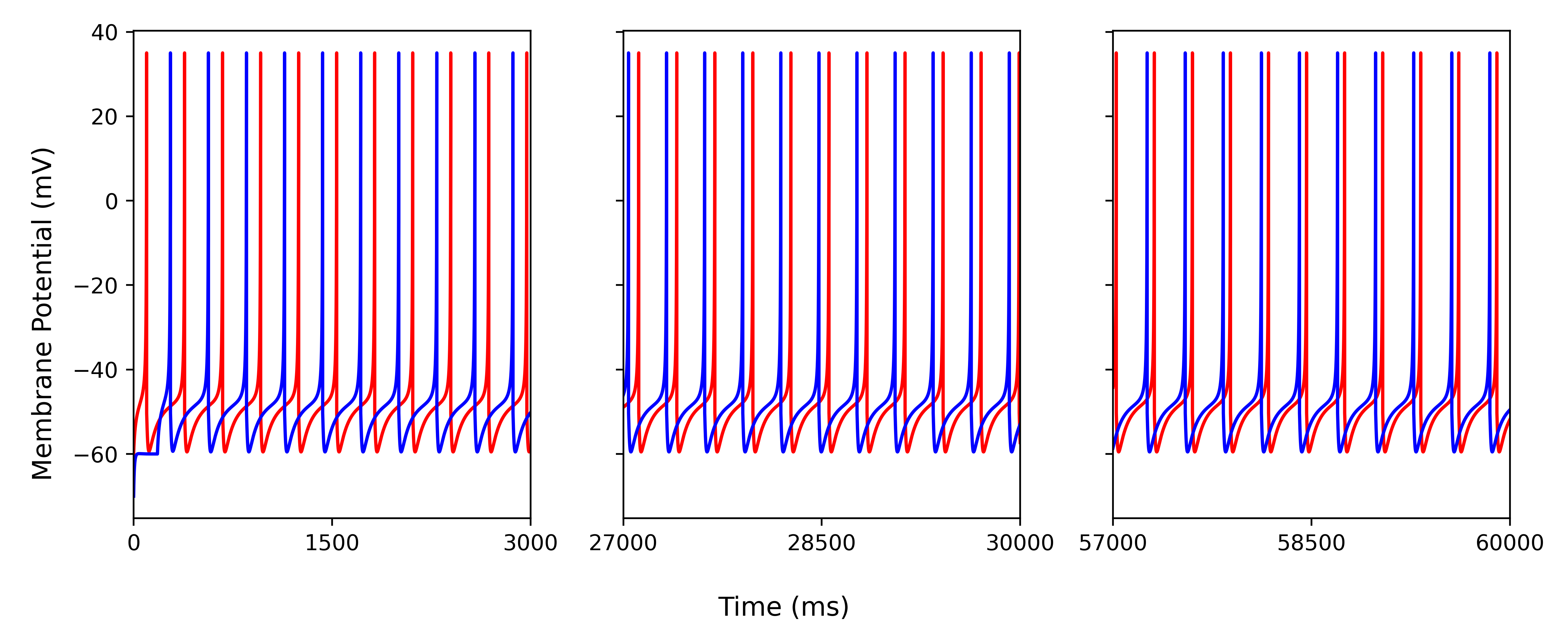}}
\caption{Numerical simulations for two coupled neurons with $b$ and $I_{app}$ as shown, $\beta=0.8$ and $g_{\rm syn}=0.01$.  All other parameters are as given in Table~\ref{tab:rs_parameters}. }
\label{fig:simulations}
\end{center}
\end{figure}

We carried out numerical simulations of the full model \eqref{eq:mod1}-\eqref{eq:modreset} as follows.
We choose the same $\Ia$ and $\beta$ as for the interaction function calculation,  and $g_{\text{syn}} = 0.01$ to enforce weak coupling. We inject one neuron with a constant current $I_{\text{app}}(t) = I_\text{app}$ and inject the other neuron with a step current $I_{\text{app}}(t - \alpha) = I_{\text{app}}$ for $t \geq \alpha$ and $0$ otherwise. The parameter $\alpha$ allows us to vary the initial phase difference between the neurons. 
All numerical simulation results agreed with the phase model predictions.
Some example simulations are shown in Fig.~\ref{fig:simulations}.

\section{Summary and Conclusions}

We used phase model analysis to study the synchronization properties of Izhikevich model neurons, focussing on the effect of the parameter associated with the conductance of the adaptation current, $b$. Our results show how the synchronization is

We reviewed analysis that shows in  the parameter space consisting of $b$ and the applied current, $\Ia$, there is a Bogdanov-Takens bifurcation in the smooth subsystem of the model. As $b$ is increased through the BT value with the the loss of stability of the equilibrium point changes from a saddle-node bifurcation  and to a subcritical Hopf. There are no stable limit cycles associated with the smooth subsystem, however the discontinuous resets in the full model give rise to large amplitude, stable discontinuous limit cycles, which represent spiking (action potentials) in the model. We observed that these discontinuous limit cycles arise in two different ways: a discontinuous saddle-node on an invariant circle bifurcation or a discontinuous saddle-homoclinic bifurcation. This latter bifurcation was also noted in Tamura et al.~\cite{tamura2009bifurcation}, who used a Poincar\'e map approach to study the bifurcations of these limit cycles in the same model with a different parameter set. 
Thus the onset of spiking in the model occurred in three ways. For $b$ small enough, spiking onset occurs via a discontinuous saddle-node on an invariant circle. For larger $b$ but less than the BT point, it occurs when a saddle-node eliminates the stable resting equilibrium. For $b$ greater than the BT point, spiking onset occurs when a
subcritical Hopf bifurcation destabilizes the stable resting equilibrium.  This suggests the presence of a second codimension two bifurcation point where the saddle-node and saddle-homoclinic occur simultaneously. Using a numerical approach we estimated the value of this bifurcation for a particular set of parameters. As would be expected from studies of smooth differential equations \cite{Erm96,ermentrout2010mathematical,izhikevich2007dynamical}, our calculations of F-I curves verify that the first bifurcation is associated with class I excitability while the others are associated with class II excitability. 

We next calculated the infinitesimal phase response curves (iPRC) and found some interesting results. For $\Ia$ close enough to the bifurcation leading to the onset of spiking, the iPRC is Type I for sufficiently small $b$ and Type II for sufficiently large $b$. While seemingly consistent with the results on excitability class, the transition from Type I to Type II iPRC occurs at a much lower value of $b$ than the transition from Class I to Class II excitability. Thus there is a significant range of values of $b$ where the model exhibits Class I excitability but has a Type II iPRC. 
Further, as $\Ia$ increases away from the spiking onset bifurcation, the negative portion of the iPRC became smaller. 

We then investigated how the parameter $b$ affects the synchronization of a pair of excitatory coupled neurons. For all parameter values we considered, the phase model predicts the synchronous solution is stable. This is consistent with results for smooth conductance-based models that adaptation currents tend to synchronize neurons connected with excitatory synapses \cite{ermentrout2001effects}. However, for a range of parameter values where the system exhibits a type II iPRC, the phase model predicts stable anti-phase solutions coexist with the synchronous solution.
For values of $\Ia$ close to the spiking onset bifurcation point, Type II iPRC always corresponded to a stable anti-phase solution in the phase model. As $\Ia$ increased away from this point, the stable anti-phase solution in the phase model is lost although the iPRC remains Type II. 
This is not too surprising, given the change in the iPRC described above. For large enough $\Ia$ while the iPRC is technically Type II, the negative portion is so small that it does contribute enough to cause stability of the anti-phase solution.
In all cases, the predictions of the phase model is consistent with numerical simulations of the full model for two neurons coupled with excitatory synapses. 

The range of parameter values where the model exhibits class I excitability and Type II PRC seems surprising since for conductance-based models, it has been shown 
that Class I oscillators often have Type I PRCs \cite{Erm96}. There are two possible explanations for this. The first is that near the transition point from Class I to Class II the discontinuous homoclinic bifurcation creating the stable limit cycle and the saddle-node bifurcation eliminating the stable equilibrium point occur close together, making the 
point difficult to resolve numerically, either through bifurcation analysis of the model or calculation of FI curves.  
The second is that the analysis in \cite{Erm96} is based on reducing the model to the normal form for the SNIC bifurcation, (the theta model). An essential assumption of this approach is that at the SNIC there is smooth homoclinic curve from the saddle-node point to itself. In the Izhikevich model, this is replaced by a discontinuous homoclinic curve, thus there is no reason to a priori assume that the analysis applies. 
 
It has been observed in some conductance-based models with adaptation currents, that the switch in the phase model and synchronization of coupled neurons occurs after the BT point where the model switches from Class I to Class II \cite{ermentrout2001effects,park2016weakly,al2021m}. 
In contrast, we found that this switch occurs before the BT point and is associated with a (discontinuous) saddle-node homoclinic bifurcation. 
A similar(smooth) bifurcation sequence has been found by varying the leak conductance in conductance-based models \cite{kirst2015fundamental} and the cubic normal form for the BT bifurcation \cite{pereira2015bogdanov}. Thus, in principle this observation should not be surprising, but perhaps has not been widely noted in conductance-based models.

The coexistence of stable synchronous and anti-phase solutions is only observed close to the bifurcation point in both the phase model and simulations. While it is known that frequency can affect synchronization \cite{lewis2004dendritic,saraga2004location}, this does not appear to be a frequency effect in our model.  We observe that the frequency could vary more strongly between parameter sets with different $b$ and $\Ia$ values exhibiting bistability than parameter sets with the same $b$ and different $\Ia$ where one exhibits bistability the other does not. 

Ladenbauer et al.~studied synchronization in the adaptive exponential (ADEx) model  \cite{ladenbauer2012impact}. This model has a very similar structure to the Izhikevich model; indeed the adaptation current identical, the main difference between the two models is the form of the nonlinearity in the voltage equation \cite{brette2005adaptive,Gerstner_2014_neuronal}.
Ladenbauer et al.~also observed a switch 
from PRC Type I to II as the conductance of the adaptation current increased.  
They indicate that the switch of excitability from Class I to Class II and the iPRC from Type I to Type II occurs at the BT point. However, their plots of the iPRC switch seem to indicate that it actually occurs before, although close to, the BT point. Thus it seems their model exhibits a similar mismatch of excitability class and iPRC type to what we observe. It is possible that their model also exhibits a discontinuous saddle-node homoclinic bifurcation. They also observed a switch of iPRC type from Type II to Type I away from the onset of spiking, fairly far from the bifurcation point. Overall, the variation of the iPRC type with the parameters of the adaptation current is similar between the two models.  
Further, both models have large ranges of parameters where the adaptation current promotes synchronization between two neurons with excitatory coupling. This similar synchronization effect reproduces results in conductance-based models, where currents such as $I_{AHP}$ and the muscarinic slow potassium current, $I_M$ have been shown to have a similar synchronization effects for different neural models \cite{ermentrout2001effects,stiefel2008cholinergic,Al-Darabsah2021-ze}. The main difference between the models, is that iPRC type seems to be a stronger predictor of synchronization for the AdEx model considered by Ladenbauer et al.~than for the Izhikevich model we studied.

\bibliographystyle{elsarticle-num} 
\bibliography{refs}
\appendix
\eject
\section{Supplementary Material}\label{ap:supmat}

\begin{figure}[h!]
\begin{subfigure}{0.5\textwidth}
 \centering
 \caption{$b=2,\ \Ia = 91.3$}
\includegraphics[width=\textwidth]{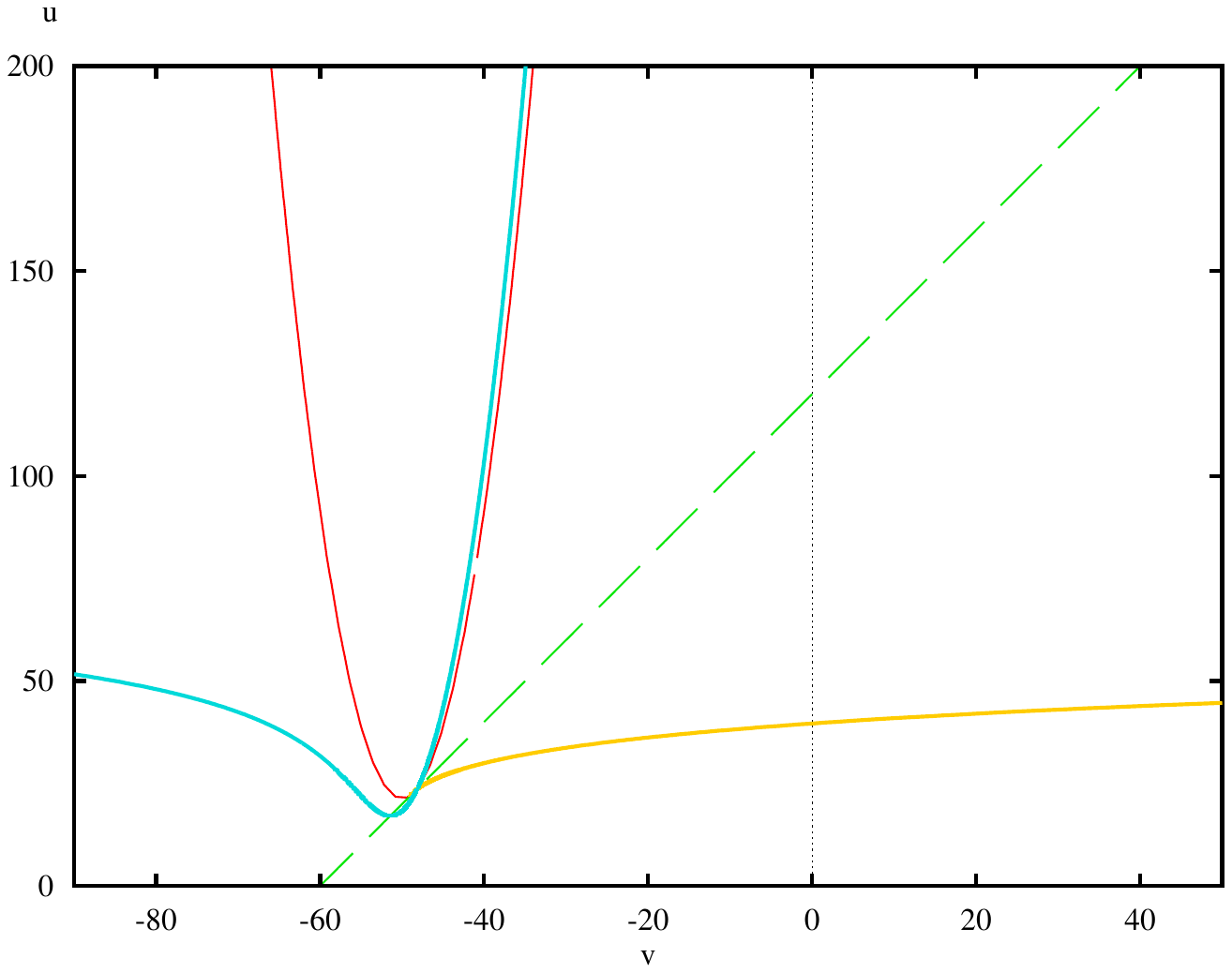}
\end{subfigure}
\begin{subfigure}{0.5\textwidth}
 \centering
 \caption{$b=2,\ \Ia = 91.34$}
\includegraphics[width=\textwidth]{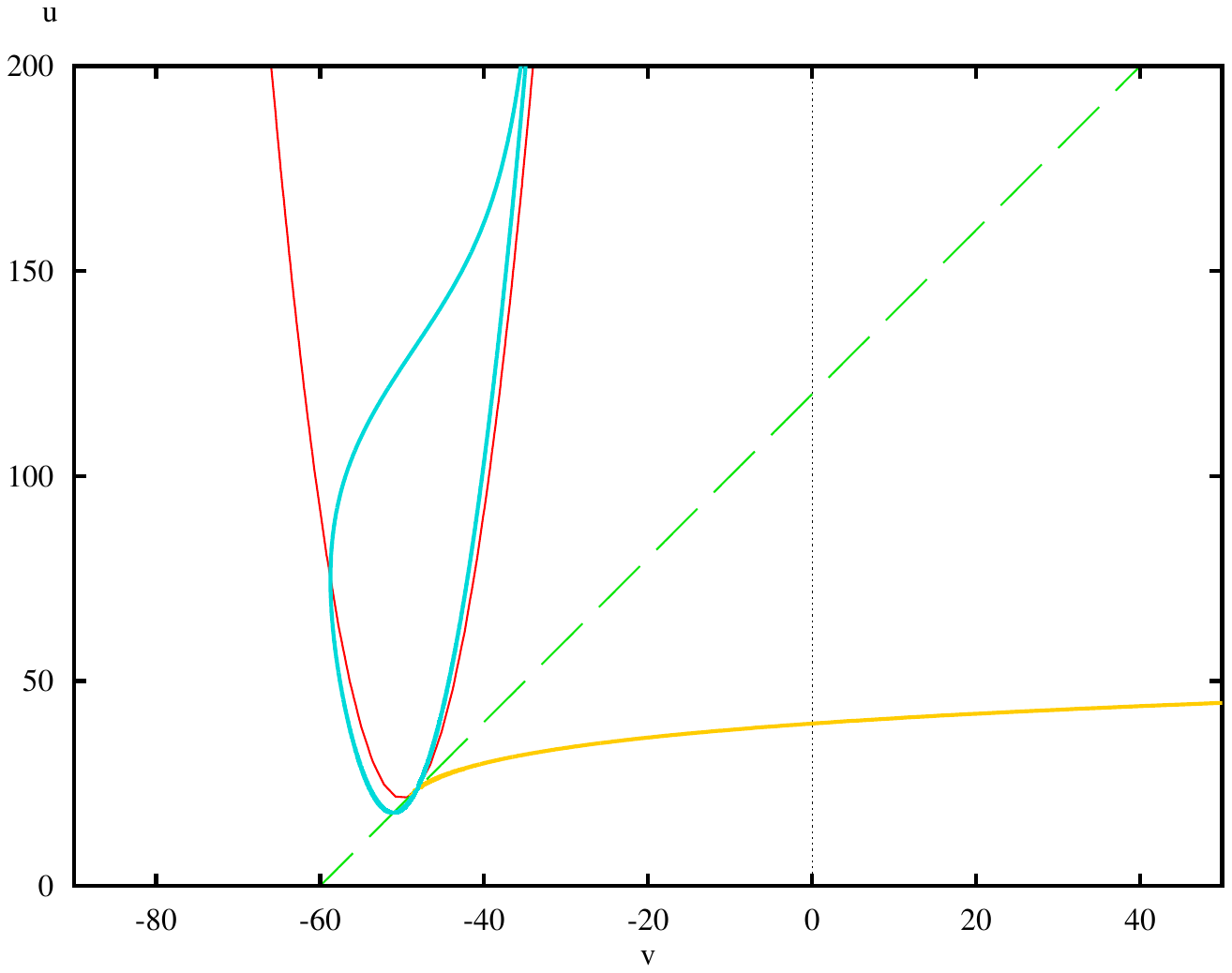}
\end{subfigure}
\caption{Phaseplane of the smooth subsystem \eqref{eq:mod1}-\eqref{eq:mod2}, showing large change to stable manifold of saddle point (cyan). Other curves are: unstable manifold of saddle  point (yellow), $v$-nullcline (red) and , $u$-nullcline (green). Initial conditions above the stable manifold give solutions that converge to the stable node. Initial conditions below the stable manifold give solutions that diverge. Parameter values are as given in Table~\Ref{tab:rs_parameters} except as shown. Generated with XPPAUT \cite{ermentrout_simulating_2002}.}
\label{fig:manifolds}
\end{figure}
\begin{table}[h!]
\begin{center}
\begin{tabular}{|c|c|c|c|c|c|}
\hline
$b$ & $I_{\rm bif}$ & Excitability & $\Ia$  & iPRC & Phase model prediction \\
\hline
$-2$ & 51.43 & Class I& $51.48$ & Type I  &  anti-phase unstable \\
\hline 
$-1.5$ &  55.80 & Class I & $55.83$ & Type I  & anti-phase unstable \\
\hline 
$-1.0$ & 60.36 & Class I & $60.38$ & Type I  &  anti-phase unstable \\
\hline 
$-0.5$ & 65.09 &  Class I & $65.11$ & Type I  &anti-phase unstable \\
\hline 
$0$ & 70.00 & Class I &  $70.01$ & Type I  & anti-phase unstable \\
\hline 
$0.5$  & 75.09 & Class I & $75.1$ & Type I  & anti-phase unstable \\
\hline 
\multirow{2}{*}{$0.6$} & \multirow{2}{*}{76.13} &  \multirow{2}{*}{Class I} & $76.14$ & Type II & anti-phase stable \\
\cline{4-6} 
&  &  &$77.00$ & Type II & anti-phase unstable \\
\hline 
\multirow{2}{*}{$1.0$} & \multirow{2}{*}{80.36} & \multirow{2}{*}{Class I} & $80.36$ & Type II & anti-phase  stable\\
\cline{4-6} 
&  & & $81.16$ & Type II & anti-phase  unstable\\
\hline 
\multirow{2}{*}{$2.0$} & \multirow{2}{*}{91.43} & \multirow{2}{*}{Class II} & $91.35$  & Type II & anti-phase stable\\
\cline{4-6} 
 && &  $92.34$ & Type II &  anti-phase unstable \\
 
\hline 
\multirow{2}{*}{$3.0$} & \multirow{2}{*}{103.21} & \multirow{2}{*}{Class II} & $102.44$ & Type II & anti-phase stable \\
\cline{4-6} 
 &  &   & $103.21$ & Type II & anti-phase unstable\\
\hline 
\multirow{2}{*}{$4.0$} & \multirow{2}{*}{115.36} & \multirow{2}{*}{Class II} & $113.51$ & Type II & anti-phase stable \\
\cline{4-6} 
& & & $115.36$& Type II  & anti-phase unstable \\
\hline
\multirow{2}{*}{$5.0$} & \multirow{2}{*}{127.50} & \multirow{2}{*}{Class II }& $124.63$& Type II  & anti-phase stable \\
\cline{4-6}
& & & $124.95$& Type II  & anti-phase unstable \\
\hline
\end{tabular}
\end{center}
\caption{Phase model predictions about stability of anti-phase solution. For these parameter values the in-phase solution is always stable. $I_{\rm bif}$ is the bifurcation value such that for $\Ia>I_{\rm bif}$ there no longer is a stable equilibrium
in the single neuron model. For $b<3$ this is a saddle-node bifurcation.
For $b> 3$ this is a subcritical Andronov-Hopf bifurcation.
}
\label{tab:Hfunresults}
\end{table}
\end{document}